\documentclass[final,5p,times,twocolumn]{elsarticle}

\usepackage{graphicx}
\usepackage{amssymb,amsmath,bm}
\usepackage{amsthm}
\usepackage{txfonts}
\usepackage{color}
\usepackage{subfig}
\usepackage[normalem]{ulem}
\usepackage{soul,xcolor}

\PassOptionsToPackage{hyphens}{url}\usepackage{hyperref}

\journal{Journal of Non-Newtonian Fluid Mechanics}

\begin{document}
\setstcolor{red}

\begin{frontmatter}


\title{Peeling of linearly elastic sheets using complex fluids at low Reynolds numbers}



\author[purdue]{Anirudh Venkatesh}

\author[purdue]{Vishal Anand}

\author[purdue]{Vivek Narsimhan\corref{corrauth}}
\ead{vnarsim@purdue.edu}
\ead[url]{https://viveknarsimhan.wixsite.com/website}

\cortext[corrauth]{Author to whom correspondence should be addressed.}

\address[purdue]{Davidson School of Chemical Engineering, Purdue University, West Lafayette, Indiana 47907, USA}

\begin{abstract}
We investigate the transient, fluid structure interaction (FSI) of a non-Newtonian fluid peeling two Hookean sheets at low Reynolds numbers ($Re \ll 1$).  Two different non-Newtonian fluids are considered – a simplified Phan-Thien-Tanner (sPTT) model, and an inelastic fluid with shear thinning viscosity (generalized Newtonian fluid). In the limit of small gap between the sheets, we invoke a lubrication approximation and numerically solve for the gap height between the two sheets during the start-up of a pressure-controlled flow.  What we observe is that for an impulse pressure applied to the sheet inlet, the peeling front moves diffusively ($x_f \sim t^{1/2}$) toward the end of the sheet when the fluid is Newtonian. However, when one examines a complex fluid with shear thinning, the propagation front moves sub-diffusively in time ($x_f \sim  t^{\alpha}, \alpha < 0.5$), but ultimately reaches the end faster due to an order of magnitude larger prefactor for the propagation speed.  We provide scaling analyses and similarity solutions to delineate several regimes of peeling based on the sheet elasticity, flow Weissenberg number (for sPTT fluid), and shear thinning exponent (for generalized Newtonian fluid).  To conclude, this study aims to afford to the experimentalist a system of knowledge to  \textit{apriori} delineate the peeling characteristics of a certain class of complex fluids.

\end{abstract}



\end{frontmatter}






















\section{Introduction}

Microfluidics- the scientific study of flows at the microscale and their concomitant phenomena \cite{whitesidenature} - has managed to hold and sustain the interests of researchers \cite{bruus2007theoretical}, engineers \cite{micro_biotech, whitesides_dig_developing} , and entrepreneurs \cite{StrohmeierO2015Cmpa} for some decades. Even though the governing principles of fluid mechanics at the macroscale are applicable unchanged at the microscale, the high surface area to volume ratio of microscale devices \cite{wereleybook} perpetuates a cascade of interesting, even unexpected, physical phenomena including, slip at the interface \cite{shu2017fluid}, flows actuated by the `weak force' of surface tension \cite{CHAKRABORTY2007175}, electrohydrodynamics \cite{ZENG2001107}, inertialess compressibility \cite{anand2020transient,gat_frankel_weihs_2008} among others. For instance, and relevant to the context of this paper, the softness of  the material (e.g., PDMS) used in microfluidic conduits affects the fluid flow considerably \cite{christov2021soft} and has been leveraged in applications such as actuators\cite{jeon2002}, valves\cite{unger2000monolithic}, bio-mimetic micro devices,  \cite{Lotters_1997} and soft robotics\cite{elbaz2018gat}. Inside these conduits, a slowly-moving viscous fluid  exerts pressure on the soft walls constituted of PDMS. This deformation alters the domain of fluid flow thereby influencing the flow field non-trivially. This two-way coupled interaction between the fluid flow and the structure is analysed under the umbrella term of fluid structure interactions (FSI).  \cite{christov2021soft,dupratestone2016}. 

One of the earliest efforts that brought the idea of a deformable surface in microfluidics forward was by Gervais et al.\cite{gervais2006} where they conducted experiments to delineate the relationship between flow rate and pressure drop for a channel with a deformable top wall. They coupled the local hydrodynamic pressure and the local deformation using an empirically derived fitting parameter. Being empirically derived, their model could not $\textit{a priori}$ predict the deformation and the corresponding relationship between flow rate and pressure drop for the deformable channel. Later, Kiran Raj et al.\cite{kiranraj2017} presented a better estimate of the fitting parameter mentioned by Gervais et al.\cite{gervais2006} by using a thick plate theory. However, the empiricism and lack of completeness in the theory posed a drawback to predict these relations $\textit{a priori}$. Christov et al.\cite{christov2018} presented the first detailed theory that used a two-way coupled FSI model-- they did not assume any fitting relationship between the pressure and the deformation in the channel wall. This study helped in explaining all the experimental observations from the past deformable channel studies \cite{gervais2006,hardy2009,seker2009,cheung2012,rajsen2016,kiranraj2017}. The theory given by Christov et  al. was limited to steady state problems, and it did not deal with the transient problem.  The transient counterpart to the steady state problem of Christov et al.\cite{christov2018} was solved by Martinez-Calvo et al.\cite{martinez2020start} when they studied the start-up flow of a Newtonian fluid in a deformable micro-channel. They provided a characteristic time scale for the fluid-structure interactions that depended on the geometry and the material of the channel walls. In all of these studies, the discussion was limited to the inflation of micro channels with clamped walls. An in-depth understanding of the dynamics of sheets with free side walls getting peeled off a surface is much less understood at a micro scale.  In relation to free sheet peeling, Hosoi and Mahadevan \cite{hosoi2004peeling} studied the time taken for a sheet to peel off from a surface under the influence of the flow of a thin layer of glycerine competing against gravity and the attractive van Der Waals forces, albeit at a macro scale.


Beyond the realm of microfluidics, transient FSI has revealed itself in several other domains in engineering as well as in nature. In geophysics,  the process of hydraulic fracturing, which is employed to enhance the recovery of oil and gas  in reservoirs, is contingent upon the physics of transient FSI \cite{ball2018,Fracturing_1}.  Closely related to hydraulic fracturing by way of the underlying physics, is the process of peeling, wherein the low Re transient FSI is leveraged to separate layers of soft materials, which may or may not be previously adhered to each other. The emerging area of soft robotics \cite{anand2020transient,matia2015dynamics} provides another fascinating platform for researchers and engineers to employ transient FSI, to push the boundaries of human endeavor.  In biological flows, on the other hand, transient FSI has been investigated in the realms of both  pulmonary systems (airways) \cite{grotberg1994pulmonary} and cardiovascular system (hemodynamics) \cite{Grotberg_Jensen}. In the former, the structure (tube) is frequently modeled as collapsible, where the radius /area of the tube decreases as a result of FSI due to negative transmural pressure \cite{heil_1997}. In the latter,  however, inflation is the norm, and the increase in radius of the arteries and capillaries due to pulsatile flow of blood conveyed by them is a phenomena widely studied and extensively researched for several decades \cite{Womersley_1957}. More recently, the focus in biological flows FSI has shifted towards the investigation of brain aneurysms where the blood vessel inside the brain softens itself and develops a bulge locally to relieve the high blood pressure \cite{boster2022challenges}.

Newtonian rheology is an idealization, and deviations from this idealization abound both in nature (biological fluids \cite{bird1987dynamics}) and in laboratory (high molecular weight polymers \cite{chhabra2011non}). Such deviations include, but are not limited to, shear dependent viscosity, nonlinear relationship between stress and strain rate, memory of the stress field, among others. One of the first forays into the FSI of non-Newtonian fluids was made by the Anand et al. \cite{anand2019non}, who investigated shear thinning fluids inside both thin and thick walled structures. They deduced that compared to the case of the Newtonian fluids, the shear thinning fluids show smaller deformation, whilst, counter-intuitively, the thick-walled channel show higher deformation compared to the thin-walled case. This study has been further extended to make use of even more realistic models that incorporate viscoelasticity \cite{arzolabautista2021}.

The fluid structure interaction in the creeping flow regime has also been analysed from the point of view of fluid instabilities. Inertialess flow of a Newtonian fluid in a rigid channel is stable because there are no time dependent terms; the introduction of softness in the walls alters this \textit{status quo}. In general, wall viscosity stabilizes the flow, while wall elasticity destabilizes it \cite{kumaran_1994,kumaran_1995}. On the other hand, the non Newtonian rheology of the fluid further complicates the situation \cite{ROBERTSPower}. For the plane Couette flow of a viscoelastic fluid past a linearly elastic solid, the system becomes unconditionally stable beyond a critical relaxation time of the fluid: no amount of solid elasticity can make the flow unstable if the fluid elasticity is beyond a critical value \cite{SHANKARElastic}. For the plane Couette flow of a shear thinning fluid, shear thinning has a stabilizing effect if the ratio of solid to fluid thickness is small, while the converse is true for large thickness ratios \cite{ROBERTSPower}. Recently, research into instability of creeping flow FSI has forayed into the domain of finite (but still low) Reynolds numbers \cite{WangChristovFinite}. 

To summarize, a bird's eye prospect of the extant literature pertaining to microscale FSI reveals that we are at the foot of an interesting cusp. On one hand, a fair amount of ground has been covered in Newtonian FSI research, where both steady state and transient problems across a wide array of geometries- thin/thick-walled channels, peeled sheets- have been investigated with aplomb. On the other hand, research in non-Newtonian FSI  has lagged wherein only a minutiae of steady state problems fluid instabilities have been analyzed in simple geometries. Therefore in this paper, we attempt to plug this knowledge gap in non Newtonian FSI literature and answer the following questions: \textit{What are the transient characteristics of peeling actuated by non-Newtonian fluids ? How do non-Newtonian attributes such as shear thinning and viscoelasticity affect the deformation, rate of deformation, rate of front propagation and pressure drop during the peeling of an elastic sheet? What are the quantitative scales of these dependent variables which may be deduced a a priori and be made  available in the service of the experimentalist  and designer?} 

The outline of this paper is as follows.  Section \ref{GE} describes the problem setup of a non-Newtonian fluid peeling two linearly elastic sheets. This consists of the geometry, structural mechanics, and the rheological models. We discuss the properties described by the simplified Phan-Thien Tanner (sPTT) constitutive relationship, as this model captures many qualitative features observed in polymeric solutions such as shear thinning, extensional thickening, normal stress differences, and viscoelasticity. We also discuss the rheological characteristics of a classic power law model in this section. We conclude section \ref{GE} with a discussion on the physical parameters that could be realized via experiments. Section \ref{Height} discusses the numerical and similarity solutions for an sPTT fluid. For the case of startup, transient peeling under pressure controlled flow, we provide scaling relationships and similarity solutions for the peeling time and the propagation front under the limits of strong viscoelasticity and weak viscoelasticity, as well as weak and strong peeling deformations. In section \ref{lubrication_analysis:powerlaw}, we repeat the same analysis for a generalized power law fluid, which allows one to examine a much wider range of shear thinning behavior than the sPTT model. Section \ref{conclusion} gives the final takeaway messages from this study to facilitate future work in the domain.\\

\vspace{-2em}
\begin{figure}
\centering
\subfloat{
\includegraphics[width = 0.49\textwidth]{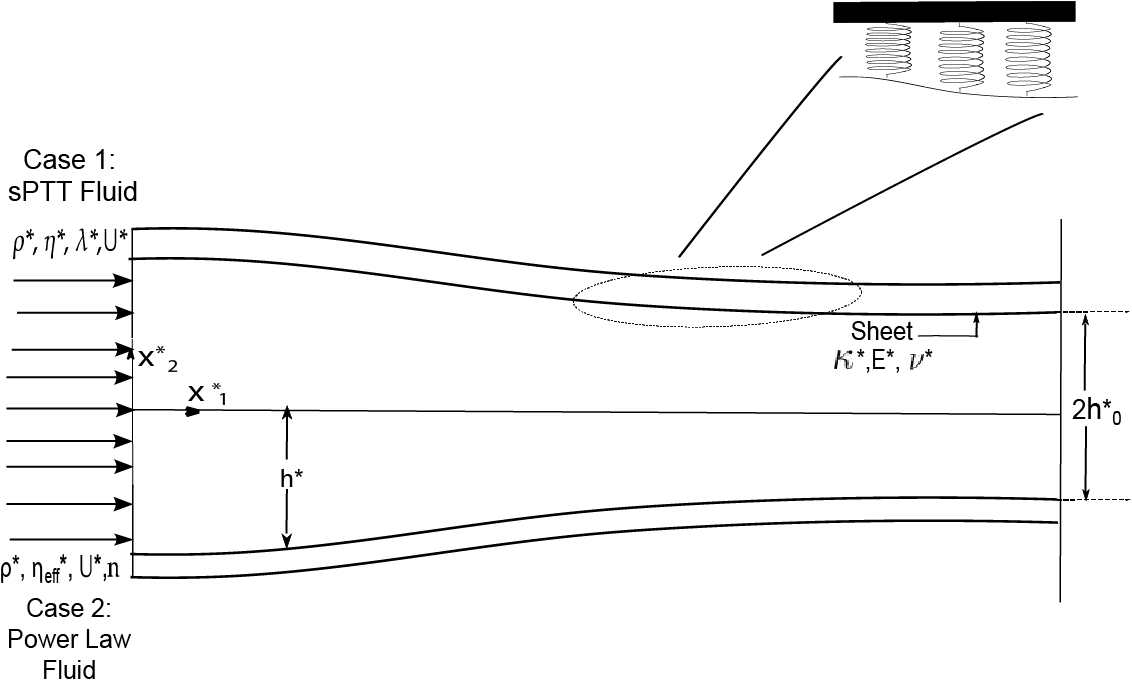}}
\vspace{0em}
\subfloat{
\includegraphics[width = 0.5\textwidth ]{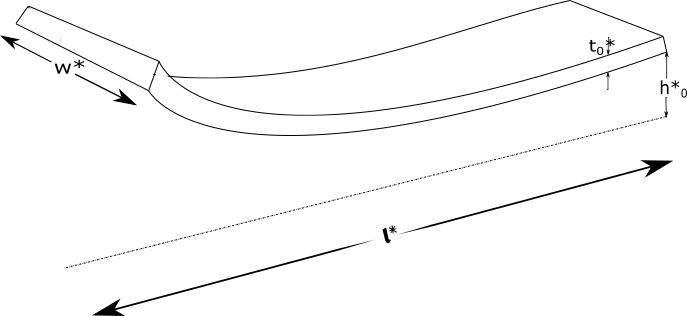}}
\vspace{1em}
\caption{  Problem schematic. (a) side view and (b) isometric view above the line of symmetry.  } 
\label{Schematic}
\end{figure}
\vspace{0em}
\section{Problem formulation}\label{GE}
\subsection{Problem geometry}\label{setup}
The schematic of the system is similar to that in Elbaz et al.\cite{elbaz2018gat} and is shown in \ref{Schematic}. In our discussion, we will denote all quantities with * as dimensional, while unstarred quantities are dimensionless. We consider an incompressible, non-Newtonian fluid with density $\rho^*$  flowing through a deformable conduit. The equations that describe the fluid flow are the Cauchy momentum equations and \st{fluid incompressibility} the equation of continuity:
\begin{equation}
\label{eq:NSeqn}
\rho^*\left( \frac{\partial \bm{u^*}}{\partial t^*} + \bm{u^*} \cdot \nabla^* \bm{u}^* \right) = -{\nabla^* p^{*}} + {\nabla^{*}}\cdot{\tau^{*}}
\end{equation}
\begin{equation}
\label{Continuity}
\bm{\nabla^{*}}\cdot\bm{u^{*}} = 0    
\end{equation}
In the above equations, $\bm{\tau}^*$ is the stress tensor for the fluid. The non-Newtonian fluids we examine are discussed in Sections \ref{sPTT_Eqnsec} and \ref{PowerLaw}.  

The coordinate system chosen for this problem is such that the $x_1^{*}$ and $x^{*}_2$ axes align with the flow and shear gradient directions, respectively.  The origin is at the center of the flow inlet, and thus the structure is symmetric about the $x_2^*$ axis.  The velocity of the fluid in the $x_{1}^{*}$ and $x_{2}^{*}$ directions is denoted by $\bm{u^{*}} = (v_{1}^{*}, v_{2}^{*})$, and the characteristic flow speed in the $x_1^*$ direction is $U^*$. The dimensional gauge pressure is given by $p^{*}$.  The initial undeformed fluid gap between the sheets  is $2h_{0}^*$ and the length of the channel along the positive $x_1^{*}$-axis is $l^{*}$.  We will consider a long, slender conduit such that the inverse channel aspect ratio $\delta = h_0^*/l^* \ll 1$.

\subsection{Structural Mechanics}\label{SM}
The sheet is treated as a linearly elastic solid with Young’s modulus $E^*$, Poisson’s ratio $\nu$, thickness $t_{0}^*$, and width $w^*$.  Here, we lump all these parameters into a single elastic stiffness $\kappa^*$ that relates the sheet’s deformation to the local pressure:
\begin{equation}
h^* = h_0^* + p^*/\kappa^*
\end{equation}
This relationship indicates that the channel walls mimic the behavior of a Hookean spring.  This statement at first may seem crude, but there are several reasons that justify this expression:

\begin{enumerate}
    \item Thin sheets undergoing peeling predominantly exhibit deformation in transverse directions, with negligible displacement in flow wise direction \cite{christov2018,anand2020,elbaz2018gat}. The same behavior is accurately portrayed by the set of linear springs, which deform only vertically and not horizontally.
    
    \item In the limit of the structure being slender, more complicated structural mechanics models like the classical plate theory, Reissner Mindlin plate theory, and Donnell shell theory yield a linear relationship between the local deformation and the hydrodynamic pressure\cite{arzolabautista2021,christov2021soft}. It has also  been shown in a recent review paper by Christov\cite{christov2021soft}, that several types of peeling geometries, including, but not limited to, rectangular channels and two dimensional channels can be modeled as Hookean springs.
\end{enumerate} 
 
 Of course, the linear relationship between the local hydrodynamic pressure load and local deformation has a long and rich history. Starting from the pioneering work of Coyle \cite{COYLE19882673}, which was later ``revived" by the twin soft lubrication papers of Stockthein and Mahadevan \cite{SMPOF,SMPRL}, the relationship was also used in  for modeling lubrication flows of Newtonian fluids in elastic cavities \cite{YinSatishPOF,YinSatish2006}
In short, the simplicity of the Hookean model coupled with its capacity to mimic a myriad of structural mechanical models makes it an obvious choice for this problem.  Table \ref{tbl:kappavalues} summarizes expressions for the stiffness parameter $\kappa^*$ for different types of plates.

An important dimensionless parameter is the FSI parameter $\beta$. It characterizes the extent of the sheet's deformation, which is the ratio between the lubrication pressure and the elastic resistance of the sheet:

\begin{equation} \label{eq:beta_defn}
    \beta = \frac{P^*_{lub}}{h^{*}_{0}\kappa^{*}} 
\end{equation}
The lubrication pressure for the different fluid models will be discussed in Sections \ref{sec_flowrate} and \ref{pressurecontrolled}. The resultant timescale that arises from the sheet's fluid-structure interaction is an FSI timescale $t_{FSI}^*$:

\begin{equation} \label{eq:FSI_timescale}
    t_{FSI}^*= \beta \frac{ l^{*}}{U^{*}} 
\end{equation}
This quantity is the time it takes for the sheet to deform to its steady profile under the limit of small deformations.

\begin{table}[h!]
\centering
  \begin{tabular}{|c|c|c|}
    \hline
     & Thin sheet $\left(\frac{t_{0}^*}{w^*} \ll 1\right)$ & Thick sheet $\left(\frac{t_{0}^*}{w^*} \sim O(1)\right)$ \\ 
    \hline
    $\kappa^*$ & $\dfrac{480E^*t_{0}^{*3}}{7w^{*4}(1-\nu^{2})}$ & $1.842455\dfrac{E^*}{(1-\nu^{2})w^*}$\\
    \hline
\end{tabular}%
  \caption{Elastic stiffness $\kappa^*$ for various width-averaged structural mechanics models where $t_{0}^*$ is thickness, $w^*$ is the width, $E^*$ is the Young's modulus, and $\nu$ is the Poisson's ratio \cite{wang2019}.}
  \vspace{-1.5em}
\label{tbl:kappavalues}
\end{table} 

\subsection{Fluid mechanics -- constitutive relationship}\label{sPTT_Section}

\subsubsection{Simple Phan Thien Tanner (sPTT model)}\label{sPTT_Eqnsec}
  The constitutive model chosen for Section 4 is the simple-Phan Thien Tanner (sPTT) model. This model utilises the Lodge-Yamamoto type of network theory to predict the effective slip of a polymeric chain network under the dynamic action of junction breakage and creation \cite{thien1977new}. At constant temperature, the model captures many qualitative features for the rheology of polymeric solutions.  For instance, under steady shear flow, the model predicts a positive first normal stress difference, as well shear thinning for both the viscosity and normal stress differences.  Under small amplitude oscillatory shear, one obtains a response characteristic of a viscoelastic liquid with one relaxation time.\cite{FERRAS201988}


The following equation describes the constitutive relationship of the sPTT fluid\cite{FERRAS201988}:

\begin{equation} \label{eq:sPTT}
f\left(\text{tr}\bm{\left(\bm{\tau^{*}}\right)}\right)\bm{\tau^{*}} + \lambda^{*} \bm{\widehat{\tau^{*}}} = 2\eta^*\bm{D^{*}}
\end{equation}
In the above equation, $f$ is a function of the trace of the stress tensor and $\bm{D}^* = \frac{1}{2}\bm{(}\bm{\nabla^{*} u^{*}} + \bm{\nabla^{*} u^{*T}}\bm{)}$ is the rate of deformation tensor. The zero shear viscosity of the polymeric fluid is $\eta^*$, the longest relaxation time is $\lambda^*$, and $\bm{\widehat{\tau^{*}}}$ is the Gordon Schowalter derivative: 
\begin{equation}\label{upper convected}
    \bm{\widehat{\tau^{*}}} = \frac{D \bm{\tau^{*}}}{Dt^{*}} - \left(\nabla\bm{u}^{*}\right)^{T}\cdot\bm{\tau^{*}} - \bm{\tau^{*}}\cdot\left(\nabla\bm{u}^{*}\right)
\end{equation}
In the above expression, $\frac{D}{Dt^*}$ is the material derivative -- i.e., $\frac{D}{Dt^*} = \frac{\partial}{\partial t^*} + \bm{u}^* \cdot \nabla^*$.

The function $f$ can take many forms as long as it satisfies the properties that $f(0) = 1$, $f$ is increasing, and $f(x) \approx x$ for $x \ll 1$.  Two of the most popular forms for $f$ are given by\cite{FERRAS201988}:
\begin{subequations}
\begin{align}
\label{eq:constitutive_exponential}
    f(\text{tr}(\bm{\tau^{*}})) & = \exp\left(\frac{\varepsilon \lambda^{*}}{\eta^*}\text{tr}(\bm{\tau^{*}})\right)\\
\label{eq:constitutive_linear}
    f(\text{tr}(\bm{\tau^{*}})) &=1+\frac{\varepsilon \lambda^{*}}{\eta^*}\text{tr}(\bm{\tau^{*}})
\end{align}
\end{subequations}

Here, $\varepsilon$ is the elongation parameter that acts as a representative quantity to understand the elongational stretching and relaxation of the polymer macro-molecules. The sPTT model given by Eq.\eqref{eq:constitutive_exponential} is known as the exponential sPTT model whereas the linear model in Eq.\eqref{eq:constitutive_linear} gives the relationship for small (limiting) values of the trace of the stress tensor.  We will consider the linear sPTT model Eq.\eqref{eq:constitutive_linear} in this paper. 

One important dimensionless quantity that arises from this model is the Weissenberg number, which is the product of the polymer relaxation time and the characteristic shear rate $\dot{\gamma}^* = U^*/h_0^*$:

\begin{equation}
    Wi \equiv \frac{\lambda^* U^*}{h_0^*}
\end{equation}

A value of $Wi \sim O(1)$ indicates that the viscoelastic contributions to the stress tensor are significant.  We will consider potentially large Weissenberg numbers $Wi \sim O(1)-O(10)$ in this paper but cases where $\delta Wi \equiv \lambda^* U^*/l^* \ll 1$ -- in other words, the polymer relaxation time is smaller than the average residence time $t^*_{conv} = l^*/U^*$ in the conduit.  In recent times, there have been studies where $\delta \sim O(10^{-4})$ and lower\cite{mehboudi2019experimental} thereby allowing this simplification. We also note that for larger values of $Wi$, the onset of viscoelastic turbulence could affect the estimated predictions \cite{KumarArdekani2022}. A more detailed stability analysis (linear or nonlinear) is needed to explain these effects \cite{larson_shaqfeh_muller_1990}.

\subsubsection{Generalized Newtonian fluid }\label{PowerLaw}

 The constitutive model chosen for Section \ref{lubrication_analysis:powerlaw}  is a generalized Newtonian fluid, an inelastic fluid that exhibits shear thinning. In its most general form, the dimensional form of the stress tensor is:

\begin{equation}
    \bm{\tau}^* = 2^n K^* \left(\sqrt{ \bm{D}^* : \bm{D}^* } \right)^{(n-1)} \bm{D}^*
\end{equation}
where $\tau^*$ is the stress tensor, $\bm{D}^*$ is rate of strain tensor, $n$ is a power law index, and $K^*$ is a consistency index.  In a simple shear flow $u_x^* = \dot{\gamma}^* y^*$, the shear stress behaves as a Newtonian fluid with an effective viscosity that changes with the shear rate -- i.e., $\tau_{xy}^* = \eta_{eff}^*\dot{\gamma}^*$, where $\eta_{eff}^* = K^* |\dot{\gamma}|^{n-1}$.  Note -- this model works well for describing the steady shear behavior of many fluids where one observes a power law scaling for viscosity versus shear rate.  Some examples include Xanthan gum and aqueous polyacrylamide solutions \cite{ZHONG2013160,Ansari2020-nz}. We also note that this model describes a wider range of shear thinning behavior than the sPTT model discussed in section \ref{sPTT_Eqnsec}, at the expense of neglecting other effects such as elasticity and normal stress differences.

\subsection{Parametric analysis and summary of dimensionless numbers}\label{ParametricAnalysis}

Tables 2 and 3 list values of the dimensionless parameters we will examine in this study, which correspond to typical values one can find in microfluidic experiments with polymeric solutions and PET and PDMS sheets\cite{mehboudi2019experimental,Dangla2010}.
The values of geometrical parameters we consider are of the order of the dimensions mentioned in the study performed by Mehboudi et al.\cite{mehboudi2019experimental} where the channel height, width, and length are $h_0^* = 2.5$ $\mu$m, $w^* = 2000$ \st{m}$\mu m$ and $l^* = 1$ cm, respectively.  The typical flow rate through such conduits is $q^* = 0.1$ $\mu$L/min, which gives the average flow speed through the channel being $U^* = q^*/(2h_0^* w^*) = 1.667 \times 10^{-4}$ m/s.  The elastic sheet that we consider is PDMS with a $E^* = 1.6$ MPa, Poisson's ratio $\nu = 0.5$, and thickness $t_{0}^* \approx 0.6$  mm, giving an elastic stiffness $\kappa^* = 1.97 \times 10^{9}$ Pa/m according to the equation for $t_{0}^{*}/w^{*} \ll 1$ given in Table \ref{tbl:kappavalues}.

We consider three possible fluids whose shear rheology fit the linear sPTT model.  The first example is an aqueous solution of polyvinylpyrolidone (PVP).  At 8 $\%$ weight and molecular weight 360 kDa, the density of the solution is $\rho^* \approx  1000$ kg/$\text{m}^3$, zero shear viscosity $\eta^* \sim O(0.1)$ Pa.s and relaxation time $\lambda^* \sim O(1)$ ms.  Similarly, a solution of actatic polystyrene (a-PS, 0.008$\%$ wt., 6 MDa) in dioctyl phthalate (DOP) exhibits a density $\rho^* \approx 1000$ kg/$\text{m}^3$, zero shear viscosity $\eta^* \sim O(0.1)$ Pa.s and relaxation time $\lambda^* \sim O(100)$ ms.  Lastly, a solution of actatic polystyrene (a-PS, 0.14 $\%$ wt., 6.9 MDa) in tricresyl phosphate (TCP) has density $\rho^* \approx 1000$ kg/$\text{m}^3$, zero shear viscosity $\eta^* \sim O(0.1)$ Pa.s and relaxation time $\lambda^* \sim O(10)$ ms.  All three examples exhibit Newtonian behavior at low shear rates, but shear thinning with a power law exponent $n \approx 1/3$ at high shear rates consistent with the linear sPTT model\cite{romeo2013,delgiudice2017}.

For power-law fluids, multiple experimental studies have been performed to analyse flow dynamics in complex processes. \cite{rajsen2016,EBAGNININ2009360,longo_difederico_chiapponi_2015}. These studies generally deal with fluids like $ 0.1 \%$ Xanthan gum (molecular weight - $10^{6}$ g/mol) and $0.1 \%$ polyethylene oxide(PEO) (molar weight-  $4 \times 10^{6}$ g/mol).

We see from the Tables \ref{tbl:deformation_integrals} and \ref{tbl:powerlaw} that the proposed geometry, elastic sheet, and fluid properties satisfy the following conditions -- (a) $\delta \ll 1$, $\delta Re \ll 1$, and $\delta Wi \ll 1$ for the sPTT fluid, where $Re=\rho^* U^* h_0^*/\eta^*$ is the channel Reynolds number that dictates the inertial contribution; (b) $\delta \ll 1$ and $\delta Re_{eff} \ll 1$ for the generalized Newtonian fluid, where $Re_{eff} = \rho^*U^*h_0^*/\eta^*_{eff}$ is the effective Reynolds number. Under these limits, one can perform a lubrication analysis to determine how non-Newtonian rheology affects the sheet's peeling front and time during startup, pressure controlled flow.  Sections \ref{Height} and \ref{lubrication_analysis:powerlaw} will discuss the lubrication model and results. 

\begin{table}[h!]
\centering
  \begin{tabular}{|p{0.1\textwidth}|p{0.1\textwidth}|p{0.1\textwidth}|p{0.1\textwidth}|}
  \hline
    Variable & Name & Definition & Order of Magnitude \\ 
    \hline
    $\delta$ & Inverse channel aspect ratio & $h_0^*/l^*$ & $O(10^{-4}-10^{-3})$\\
    \hline
    $Re$ & Reynolds number & $\rho^* U^* h_0^*/\eta^*$ & $O(10^{-2}-1)$\\
    \hline
    $Wi$ & Weissenberg number & $\lambda^* U^*/h_0^*$ & $O(1-10)$\\
    \hline
    $\varepsilon$ & Elongation parameter & -- & $O(10^{-2})$\\
    \hline
    $\beta$ & FSI parameter & $P_{lub}^*/(h_0^{*} \kappa^*)$ & $O(10^{-2} - 10)$\\
    \hline
\end{tabular}%
  \caption{Dimensionless numbers for channel geometry, channel elasticity, and sPTT fluid along with typical values for experiments.  The lubrication pressure is $P_{lub}^* = \frac{\eta^*U^*}{\delta h_0^*}$.} 
  \vspace{-1em}
\label{tbl:deformation_integrals}
\end{table}

\begin{table}[h!]
\centering
  \begin{tabular}{|p{0.1\textwidth}|p{0.1\textwidth}|p{0.1\textwidth}|p{0.1\textwidth}|}
  \hline
    Variable & Name & Definition & Order of Magnitude \\ 
    \hline
    $\delta$ & Inverse channel aspect ratio & $h_0^*/l^*$ & $O(10^{-3}-10^{-2})$\\
    \hline
    $\eta_{eff}^*$ & Effective viscosity & $K^* |\dot{\gamma^*}|^{n-1}$ & $O(10^{-3})-O(0.1)$  $Pa.s^{n}$\\
    \hline
    $n$ & Power law index & -- & $0.3-0.8$\\
    \hline
        $\beta$ & FSI Parameter & $P_{lub}^*/(h_0^{*} \kappa^*)$ & $O(10^{-2}-10)$\\
    \hline
\end{tabular}%
  \caption{Key parameters for a power-law fluid along with typical values for experiments.  The characteristic shear rate is $\dot{\gamma^*} = U^*/h_0^*$ and the lubrication pressure is $P_{lub}^* = \frac{\eta_{eff}^*U^*}{\delta h_0^*}$.}
  \vspace{-1em}
\label{tbl:powerlaw}
\end{table}

\section{ Lubrication analysis -- sPTT fluid}\label{Height}


\subsection{Lubrication Equations}

We invoke the lubrication approximation to solve for the fluid flow and the height profile for the deformable channel when the fluid satisfies the sPTT constitutive relationship.  The lubrication approximation assumes that the channel is long and slender such that $\delta \ll 1$ and $\delta Re \ll 1$.  This approximation allows us to neglect the acceleration terms (both Eulerian as well as convective) in the momentum equation, so the flow is nearly unidirectional:

\vspace{-1.5em}
\begin{equation} \label{eq:lubrication_Cauchy}
    -\nabla^* p^* + \nabla^* \cdot \bm{\tau}^* = 0
\end{equation}
Let us proceed with non-dimensionalization.  We scale the lengths in  in the $x_1^*$ and $x_2^*$ direction by $l^*$ and $h_0^*$ respectively, and the  velocity in the $x_1^*$ direction by $U^*$. Using equation of continuity, the velocity scale in the $x_2^*$ is revealed to be $\delta U^*$.  Time is scaled by the average residence time $t^*_{conv} = l^*/U^*$ in the conduit.  The pressure is scaled by a lubrication pressure scale $P^*_{lub}$, and the stresses are scaled by $\delta P^*_{lub}$. Depending on the problem at hand (pressure controlled or flowrate controlled), one of the two scales $P^*_{lub}$ or $U^*$ will be specified, with the other quantity obtained through  $P^*_{lub} = \eta^* U^*/(\delta h_0^{*})$.  Below are the definitions of the non-dimensional variables:


\begin{subequations}
\begin{align} 
    x = \frac{x_{1}^{*}}{l^*}; \quad y = \frac{x_{2}^{*}}{h_0^*}; \quad h = \frac{h^*}{h_0^*} \label{eq:dist_nondimensional} \\
    u = \frac{v_{1}^{*}}{U^*};\quad v = \frac{v_{2}^{*}}{\delta U^*}; \quad t = \frac{U^*t^{*}}{l^*} \label{eq:vel_nondimensional}\\
    p = \frac{p^{*}}{P^*_{lub}}; \quad \bm{\tau}= \frac{\bm{\tau}^{*}}{\delta P^*_{lub}}; \label{eq:stress_nondimensional}\\
    P^*_{lub} = \frac{\eta^* U^*}{\delta h_0^*} \label{eq:pressure_scale}
\end{align}
\end{subequations}

Using the above definitions, we write the Cauchy momentum equations Eq.\eqref{eq:lubrication_Cauchy} in non-dimensional form and discard terms of $O(\delta)$ or higher. The equations become:

\begin{subequations} 
\begin{align}
\label{NS_x}
    \frac{ \partial \tau_{xy}}{\partial y} = \frac{\partial p}{\partial x} \\
\label{NS_y}
        \frac{ \partial p}{ \partial y} = 0
\end{align}
\end{subequations}

To solve for the flow field in the channel, one needs to relate the shear stress $\tau_{xy}$ to the velocity field.  We proceed to inspect the linear sPTT model in Eq.\eqref{eq:sPTT}, Eq.\eqref{upper convected}, and Eq.\eqref{eq:constitutive_linear}. In dimensionless form, the $xx$, $yy$, and $xy$ components of the stress equations are written below, neglecting terms $O(\delta)$ and higher as well as terms $O(\delta Wi)$ as stated previously:

\begin{subequations}
\begin{align} 
\label{tau_{xx}}
    f(\tau_{xx}+\tau_{yy}) \tau_{xx} - 2 Wi \left[\frac{\partial u}{\partial y}\tau_{xy}\right] = 0 \\ 
\label{tau_{yy}}
     \tau_{yy} = 0 \\
\label{tau_{xy}}
     f(\tau_{xx}+\tau_{yy})\tau_{xy} -Wi\left[\frac{\partial u}{\partial y} \tau_{yy}\right] = \frac{\partial u}{\partial y}
\end{align}
\end{subequations}

Because we neglected the $O(\delta Wi)$ terms, there is no transient term in the constitutive viscoelastic relationship.  This means that one can treat the flow field as quasi-steady over timescales comparable to the residence time in the channel as well as the deformation of the channel.

On inspecting the equation for function $f$ in Eq.\eqref{eq:constitutive_linear}, one obtains:

\begin{equation} \label{eq:f_dimensionless}
    f(\tau_{xx} + \tau_{yy}) = 1+\textcolor{blue}{\varepsilon} Wi(\tau_{xx} + \tau_{yy})
\end{equation}

Furthermore, dividing Eq.\eqref{tau_{xx}} by Eq.\eqref{tau_{xy}} yields a relationship between the normal ($\tau_{xx}$) and shear stress ($\tau_{xy}$):

\begin{equation} \label{eq:normal_shear_relation}
    \tau_{xx} = 2 Wi \left(\tau_{xy}\right)^{2}
\end{equation}

The above relationship illustrates the nonlinear characteristics of the rheological model lucidly; there is nonzero normal stress present in the flow $\tau_{xx}$, even though there is no normal strain rate $\frac{\partial u}{\partial x} =0$.  This normal stress does not directly contribute a force on the channel wall -- however, it indirectly alters the pressure distribution by affecting the shear thinning of the fluid. We also remark that in the limit of $Wi \to 0$, the Newtonian limit is retrieved and the normal stresses are zero.
Substituting expressions Eq.\eqref{eq:normal_shear_relation} and Eq.\eqref{eq:f_dimensionless} into Eq.\eqref{tau_{xy}} gives the relationship between the shear stress and the velocity field:

\begin{equation}\label{eq:tau_xyDe}
    \left(1+2\varepsilon Wi^2 \tau_{xy}^2\right)\tau_{xy} = \frac{\partial u}{\partial y}
\end{equation}

The above captures some key features observed in the rheology of polymeric solutions.  At low shear rates ($Wi \ll 1$), the shear stress depends linearly on the shear rate:  $\tau_{xy} \sim \frac{\partial u}{\partial y}$.  At large shear rates ($Wi \gg 1$), the shear stress $\tau_{xy} \sim (\frac{\partial u}{\partial y})^{1/3}$.  This is equivalent to a shear thinning fluid with power-law exponent $n = \frac{1}{3}$. We also note that the shear thinning is direct consequence of nonlinearity of the rheological model. Had $f(\tau_{xx}+\tau_{yy}) = 1$, the model would have been linear and a linear relationship between shear stress and strain rate from Eq.~\eqref{tau_{xy}} been obtained with no shear thinning.

Using Eq.\eqref{eq:tau_xyDe}, we can now solve for the flow field in the channel.  We first integrate the $x$-momentum equation Eq.\eqref{NS_x} over the $y$ direction and enforce the symmetry condition at the channel center ($y = 0$).  This procedure yields $\tau_{xy} = \frac{\partial p}{\partial x} y$.  Substituting this expression into the stress relationship Eq.\eqref{eq:tau_xyDe} gives the shear rate in terms of the pressure gradient.

\begin{equation} \label{dudy_2}
\frac{\partial u}{\partial y} =     \left(\frac{\partial p}{\partial x}\right)y + 2\varepsilon Wi^{2}\left(\frac{\partial p}{\partial x}\right)^{3}y^{3}
\end{equation}
We integrate the above expression and enforce the no-slip boundary condition $u(y = h) = 0$ at the wall.  This procedure yields the velocity field in the channel:
\begin{equation} \label{ufuncform}
u = \left(\frac{\partial p}{\partial x}\right)\left(\frac{y^{2} - h^{2}}{2}\right) + \varepsilon Wi^{2}\left(\frac{\partial p}{\partial x}\right)^{3}\left(\frac{y^{4} - h^{4}}{2}\right)
\end{equation}

The above equation matches the expression given in \cite{oliveira_pinho_1999}. 
On further inspecting this velocity profile, we can make some comments. When the value of $\varepsilon Wi^{2} = 0$, the velocity profile collapses to a Newtonian fluid. Now, this could mean that either $Wi = 0$ or $\varepsilon = 0$. The former case deals with a fluid that does not have any elasticity thereby giving us a Newtonian profile. The latter case, however, pertains to an Oldroyd-B (Boger) fluid where there are no shear thinning effects. In this case, we would expect the normal stresses to not contribute heavily to the velocity distribution up to  leading order. We point out an important argument about the scaling of the stresses. In the scaling given in Eq.\eqref{eq:stress_nondimensional}, we take the scales for the stresses to be the same for all the three components. This assumption holds true for extremely shallow channels-- $\delta \sim O(10^{-4})$. Under this limit, the terms of order $\epsilon$ and higher can be neglected to give us an accurate description of the viscous stresses. However, we do note that if the slenderness becomes larger $\delta \sim O(0.01) - O(1)$, the normal stress terms would have considerable effect. In such analyses, the scaling is different for the three stress components  and one can solve for the flow profile using a regular perturbation  expansion in $\delta Wi$ using techniques such as the reciprocal theorem \cite{ahmedbiancofiore2021, boykostone2021}. 

We now determine how the channel height evolves over time.  We start with the continuity equation:

\begin{equation}
    \frac{\partial u}{\partial x} + \frac{\partial v}{\partial y} = 0
\end{equation}
and integrate it over the height of the channel.  Enforcing the kinematic boundary condition at the sheet interface $v(y=h) = \frac{\partial h}{\partial t}$ yields:

\begin{equation}\label{height_evolution}
    \frac{\partial h}{\partial t} = \frac{\partial}{\partial x}\left(\frac{h^{3}}{3}\left(\frac{\partial p}{\partial x}\right) + \frac{2\varepsilon Wi^{2}h^{5}}{5}\left(\frac{\partial p}{\partial x}\right)^{3}\right)
\end{equation}

This is the nonlinear partial differential equation that governs the height of the conduit. As a consistency check, we remark that in the limit of Weissenberg number going to zero $Wi \to 0$, which is the Newtonian limit, we retrieve the same equation which was earlier reportedly used for modeling of peeling of linearly elastic sheets conveying Newtonian fluids (see Eq.~($4.11$) in \cite{MCSPS19}) In order to close this equation, we note that the relationship between the local pressure and channel height is given by:

\begin{equation} \label{heightpressure}
    h = 1 + \beta p
\end{equation}
where $\beta$ is the FSI parameter discussed in Section \ref{SM}.  The following sections will solve the PDE for different cases and describe how non-Newtonian fluid rheology alters the start up effects for channel deformation.

\subsection{Numerical and analytical solution -- flow rate controlled case}\label{sec_flowrate}
For a fixed flowrate per unit width $Q^*$, we will set the characteristic velocity and lubrication pressure scales for the non-dimensional variables in eqns Eq.\eqref{eq:dist_nondimensional}-Eq.\eqref{eq:stress_nondimensional} as follows:

\begin{equation}
    U^* = \frac{Q^*}{2h_0^*}; \qquad P^*_{lub} = \frac{\eta^* Q^*}{2 \delta h_0^{*2}}
\end{equation}

Using Table 2, the definitions for the Weissenberg number and the FSI parameter for this problem are:

\begin{equation} \label{eq:De_beta_flowrate_control}
    Wi = \frac{\lambda^* Q^*}{2h_0^{*2}}; \qquad \beta = \frac{\eta^* Q^*}{2\delta \kappa^* h_0^{*3}}
\end{equation}

Below, we will write the differential equation for the non-dimensional pressure profile along the conduit.  In non-dimensional form, the volumetric flowrate per unit width is $q = \int_{-h}^{h} u dy$, where $u$ is the velocity profile.  By the definition of our non-dimensionalization, the flowrate is also $q = 1$.  Plugging the velocity profile in Eq.\eqref{ufuncform} and integrating over $y$, we obtain.
\begin{equation}\label{eq:flowratecontrolled}
    {q} = 1 = -\frac{2h^{3}}{3}\left(\frac{\partial p}{\partial x}\right) - \frac{4\varepsilon Wi^{2}h^{5}}{5}\left(\frac{\partial p}{\partial x}\right)^{3}
\end{equation}
The above is the nonlinear ODE for pressure for a fixed  flow rate at steady state.  If one specifies  the pressure at either the inlet or outlet, one can solve for the pressure profile in the conduit.  Furthermore, on using the relationship Eq.\eqref{heightpressure} between height and pressure, one can solve for the steady deformation profile $h(x)$.

\begin{figure}[h]
\centering
\subfloat{
  \includegraphics[width=0.5\textwidth]{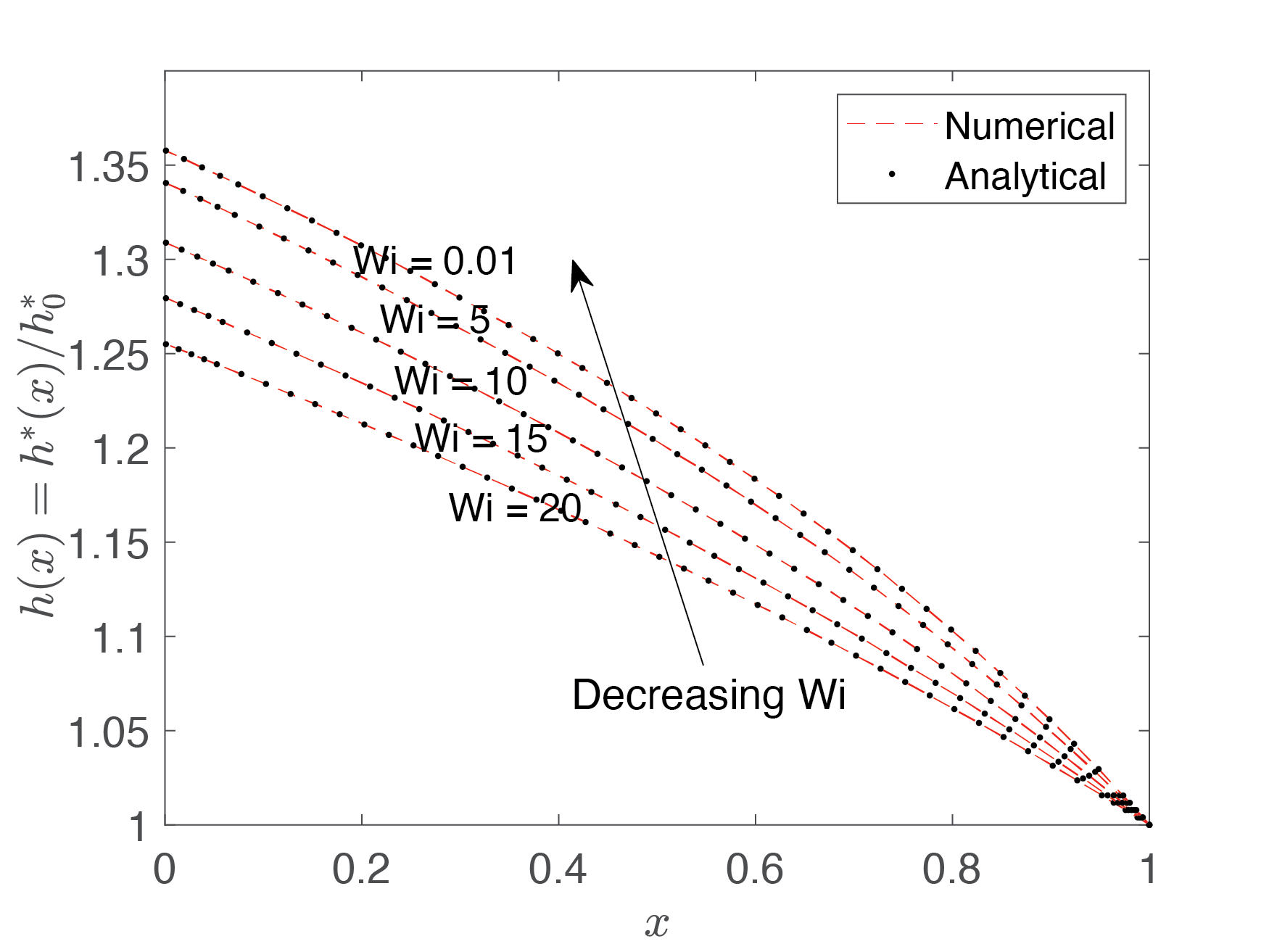}
}
\caption{Steady deformation profile of the elastic sheet  for different values of Weissenberg number ($Wi$) at fixed $\beta = 0.4$ in an sPTT fluid. The flow rate is $q = 1$. The sPTT elongation parameter $\varepsilon = 0.01$. The pressure at the outlet is kept at a normalized value of 0. Definitions of the Weissenberg number and FSI parameter are given in Eq.\eqref{eq:De_beta_flowrate_control}.}
\label{fig:flowratecontrolled}
\end{figure}

\begin{figure}
\centering
\subfloat{
  \includegraphics[width=0.5\textwidth]{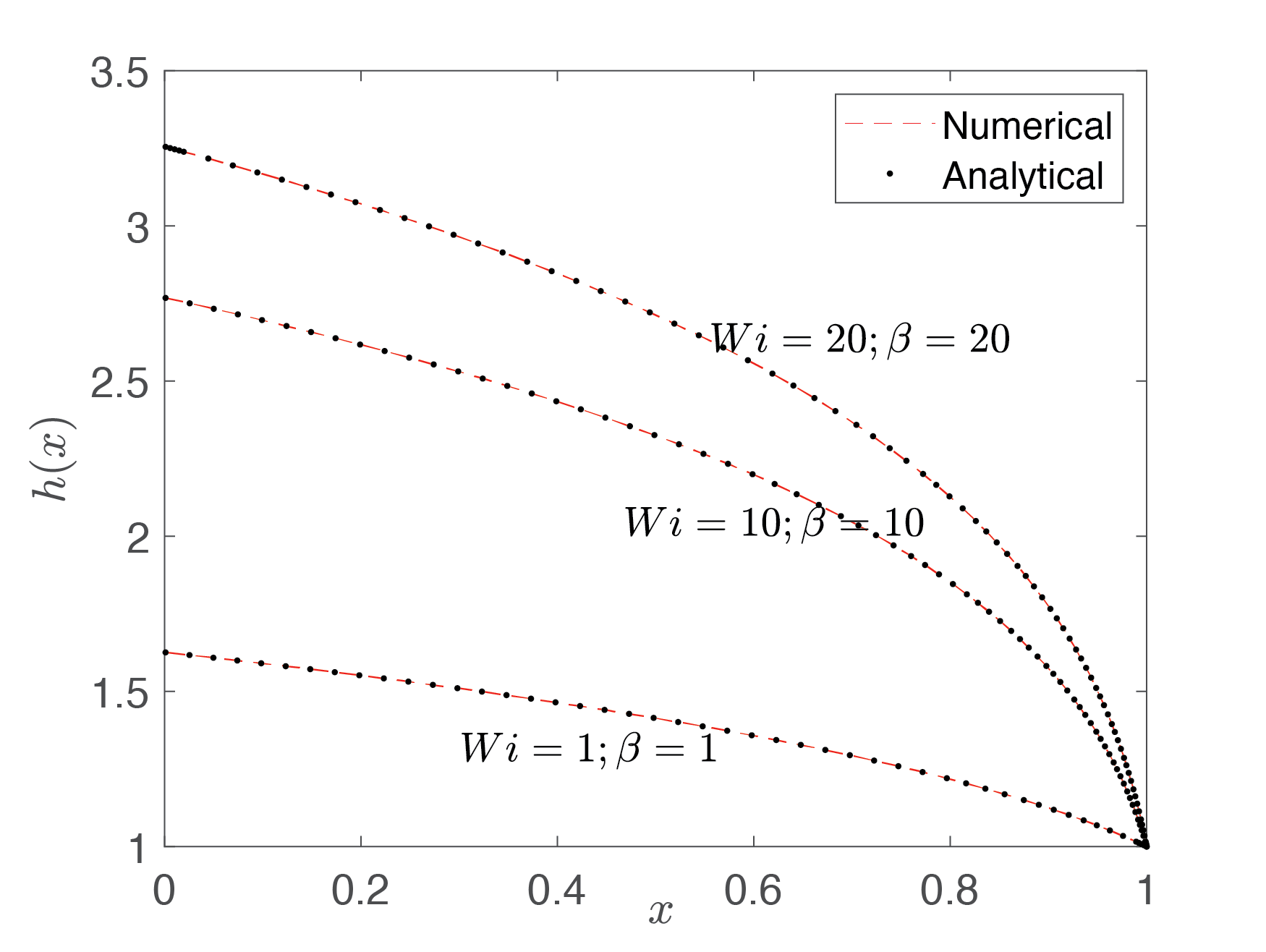}
}
\caption{Steady deformation profile of the elastic sheet in an sPTT fluid for different values of Weissenberg number ($Wi$) and fixed $\alpha = Wi/\beta = 1$. The sPTT elongation parameter $\varepsilon = 0.01$. The pressure at the outlet is kept at a normalized value of 0. Definitions of the Weissenberg number and FSI parameter are given in Eq.\eqref{eq:De_beta_flowrate_control}.}
\label{fig:flowratecontrolled_fixedalpha}
\end{figure}

Fig.\ref{fig:flowratecontrolled} and \ref{fig:flowratecontrolled_fixedalpha} plot the conduit's steady state deformation profile for different values of Weissenberg number ($Wi$) and FSI parameter ($\beta$). The curves are numerical solutions to the above ODE, while dots are the analytical solution derived in the Supporting Info (Appendix A). Fig.\ref{fig:flowratecontrolled} plots steady state deformation profiles for a fixed value of $\beta$ while varying the values of $Wi$.  This case represents a situation where one fixes the flowrate ($Q^*$), zero shear viscosity ($\eta^*$), and  sheet’s elastic properties ($\kappa^*$), but varies the fluid's relaxation time ($\lambda^*$).  As the fluid's viscoelasticity increases ($Wi$ increases) for a fixed flowrate, the deformation decreases.  Note -- a similar analysis has been discussed before for a deformable channel using an sPTT fluid and a Newtonian fluid \cite{arzolabautista2021,christov2018}. The main reason cited for this reduction in deformation is the increased shear thinning nature of the fluid.

Fig.\ref{fig:flowratecontrolled_fixedalpha} plots steady state deformation profiles when one fixes the ratio $\alpha = Wi/\beta$, but varies $Wi$.  This situation represents the case when one fixes the sheet’s elastic properties ($\kappa^*$) and fluid’s rheological properties (viscosity $\eta^*$ and relaxation time $\lambda^*$), but one varies the flowrate ($Q^*$).  For a fixed value of $\alpha=1$, we notice that increasing the Weissenberg number (proportional to flow rate) increases the maximum deformation of the channel, although the increase is nonlinear.

\subsection{Numerical solution -- pressure controlled case}\label{pressurecontrolled}






\begin{figure}
    \centering
    \subfloat
        {
            \includegraphics[width=0.4\textwidth]{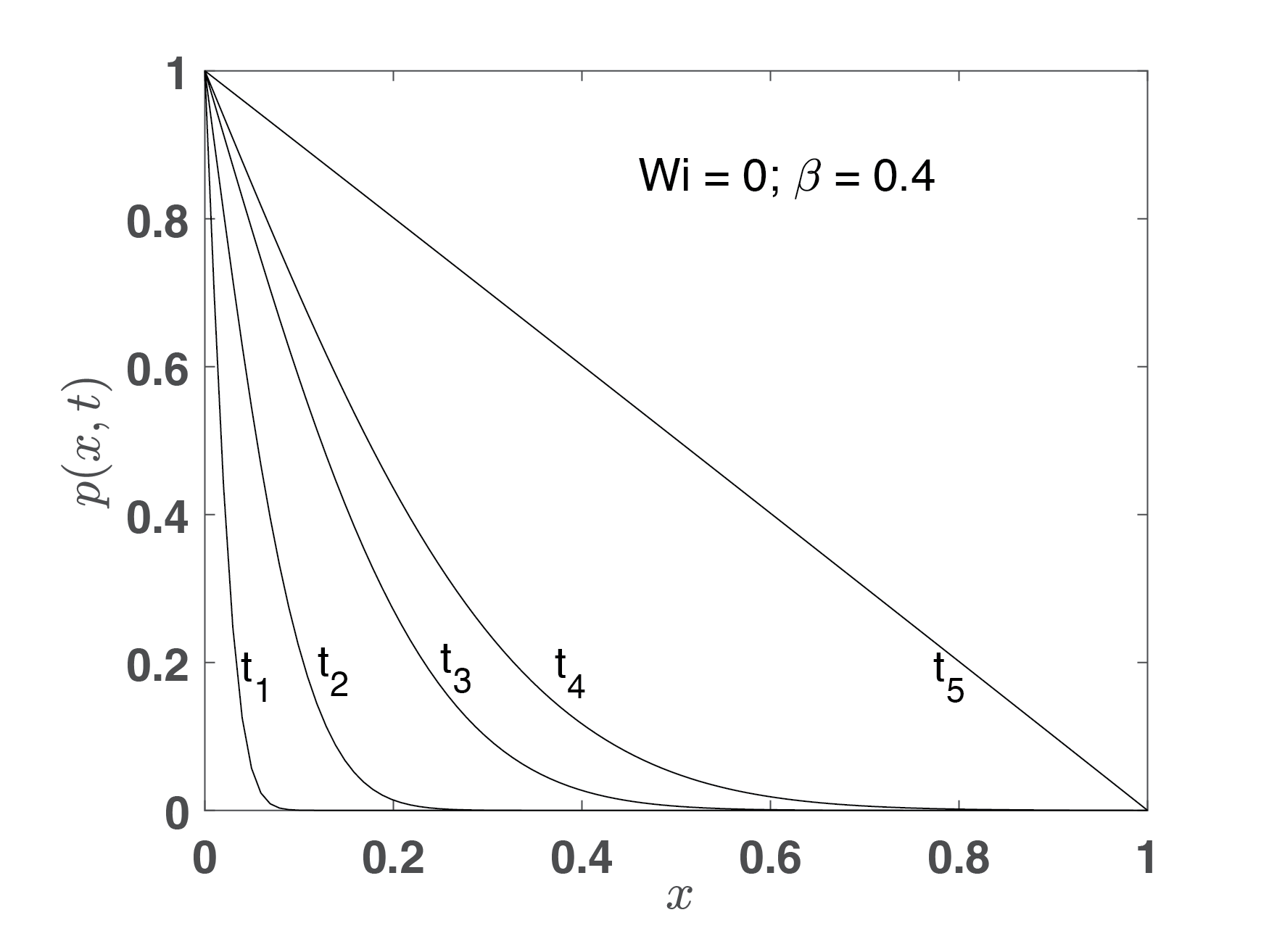}
        }
    
  \subfloat
   
        { \centering 
            \includegraphics[width=0.4\textwidth]{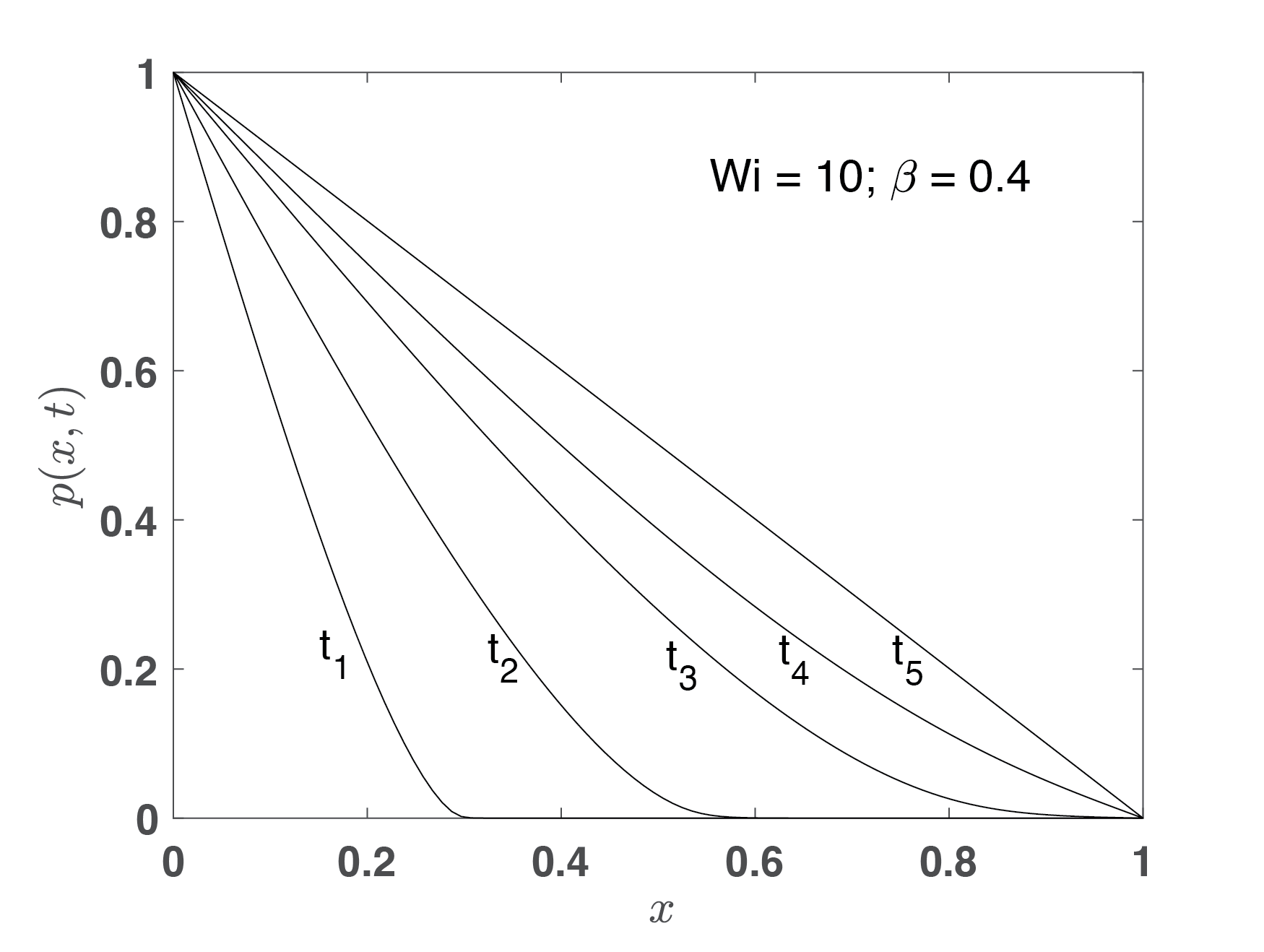}
        }
 \subfloat
    \hfill
    \centering
        {
            \includegraphics[width=0.4\textwidth]{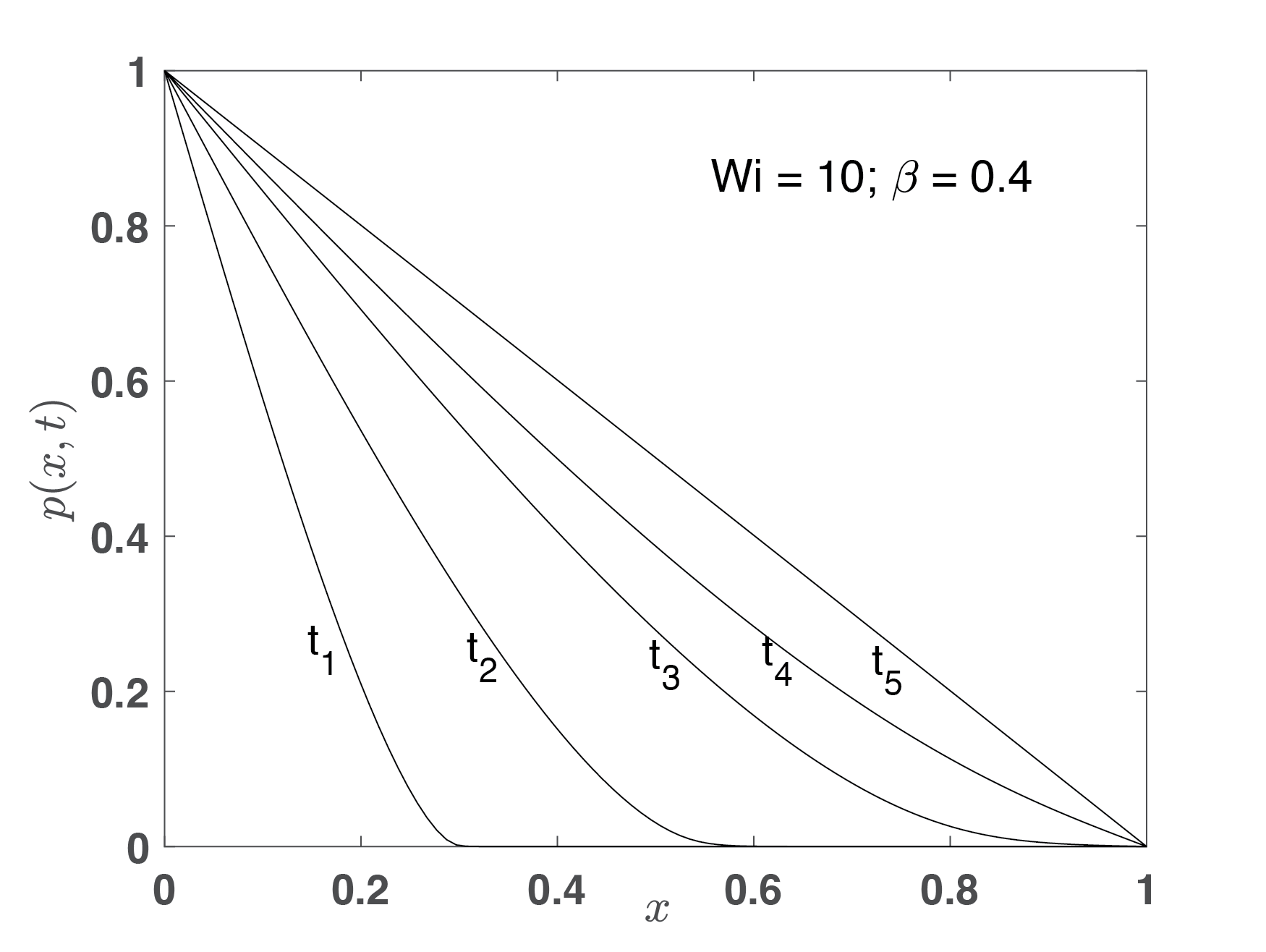}
        }
        {
            \centering
            \includegraphics[width=0.4\textwidth]{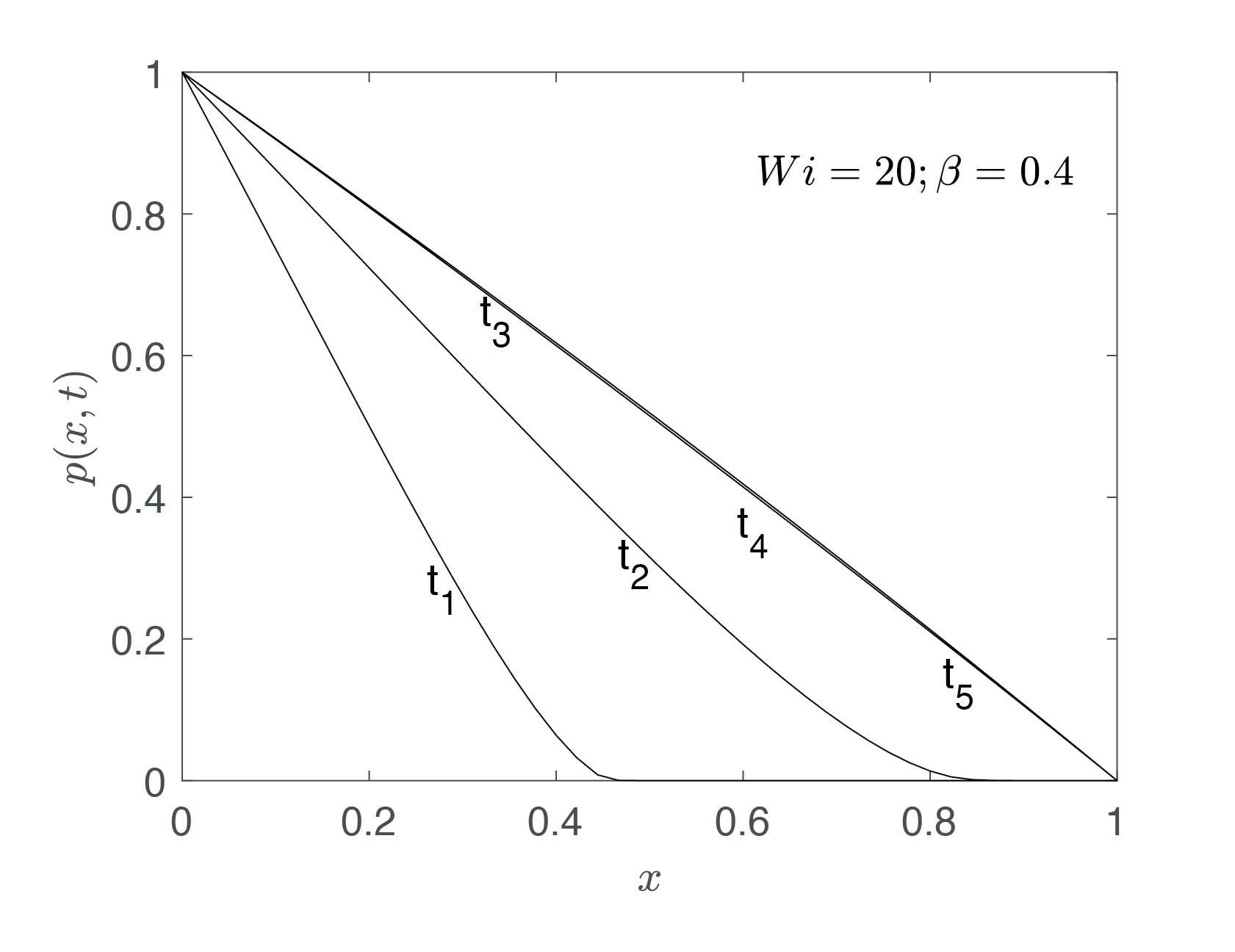}
        }
    \vspace{-1em}
    \caption{ Numerical solution for transient pressure profiles at different Weissenberg numbers $\left(Wi\right)$.  For each Weissenberg number, the pressure profiles are plotted at five different time instances $t_{1} = 0.001\beta;$ $t_{2} = 0.01\beta;$ $t_{3} = 0.05\beta;$  $t_{4} = 0.1\beta;$ $t_{5} = 2.5\beta$.  The FSI parameter is $\beta = 0.4$ for all plots. The definitions of the FSI parameter $\beta$, Weissenberg number $Wi$, and non-dimensional time $t$ are given by Eq. \eqref{defnP} and Eq.\eqref{defnDe}.} 
    \label{fig:transientplots_1}
\end{figure}

\subsubsection{Definitions}
We will examine an initially static conduit subject to a specified pressure drop $\Delta P^*$ at time $t^* = 0$. The inlet is clamped at pressure $\Delta P^*$, while the outlet is clamped at zero pressure.  The goal is to determine how the pressure profile evolves over time.  We will set the lubrication pressure and velocity scales for the nondimensional variables in eqns Eq.\eqref{eq:dist_nondimensional}-Eq.\eqref{eq:stress_nondimensional} to be:
\begin{equation}\label{defnP}
    P^*_{lub} = \Delta P^*;   \qquad U^* = \frac{\delta \Delta P^* h_0^*}{\eta^*}
\end{equation}
Using Table 2, the definitions for the Weissenberg number and the FSI parameter are:

\begin{equation}\label{defnDe}
    Wi = \frac{\delta \lambda^* \Delta P^*}{\eta^*}  \qquad \beta = \frac{\Delta P^*}{\kappa^* h_0^*}
\end{equation}
Similarly, the definition of the non-dimensional time is:
\begin{equation} \label{eq:non_dim_time_sPTT}
    t = \frac{t^* U^*}{l^*} = \frac{t^* \delta^2 \Delta P^*}{\eta^*}
\end{equation}
Given these definitions, the non-dimensional pressure distribution in the conduit will satisfy PDE Eq.\eqref{height_evolution} with the condition $h = 1 + \beta p$.  The initial pressure is $p(x,t=0)=0$ with boundary conditions $p(x=0,t) = 1$ and $p(x=1,t) = 0$. We propose two methods of solving this problem where the first is the complete numerical solution of Eq.\eqref{height_evolution} and the second is a similarity solution that aids in our understanding of the peeling rates.

\subsubsection{Numerical solution} \label{numsol}
The PDE for pressure (Eq.\eqref{height_evolution}) is a nonlinear partial differential equation with Dirichlet boundary conditions.  These class of equations can be solved using a nonlinear finite difference scheme \cite{Ghodgaonkar2019,Langtangen}.  We divide the domain $(x,t)$ into $N$ points and $M$ time steps. Each spatial step has a size $\Delta x$ and each time step has a size $\Delta t$. We use a central difference scheme in space and a backward implicit difference scheme in time. On discretizing the equations, we get $N$ nonlinear algebraic equations for the pressure values at the next time step ($p_i^{j+1}$) in terms of the pressure values at the current time step ($p_i^j$), where $i = 1,...,N$ are the indices for the spatial points and $j$ is the index for time.  These nonlinear equations are solved iteratively at each time step using Newton's method.  We typically use time steps of $\Delta t = 10^{-8} - 10^{-6}$ and grid size $\Delta x$ between $0.01$ and $0.0223$.  Our code has been tested for convergence.

Figure \ref{fig:transientplots_1} plots typical pressure profiles obtained from our numeric scheme at different values of Weissenberg number ($Wi$).  As mentioned previously, we set the initial pressure to be the static pressure $p(x,t=0) = 0$, and we specify the inlet and outlet to be at $p(x=0,t) = 1$ and $p(x=1,t) = 0$.  We observe two major trends. First, we see that the pressure profile behaves as a front that propagates towards a steady state.  Second, we see that increasing the Weissenberg number ($Wi$) hastens the time to reach steady state.  


\begin{figure}
\centering
  \includegraphics[width=0.5\textwidth]{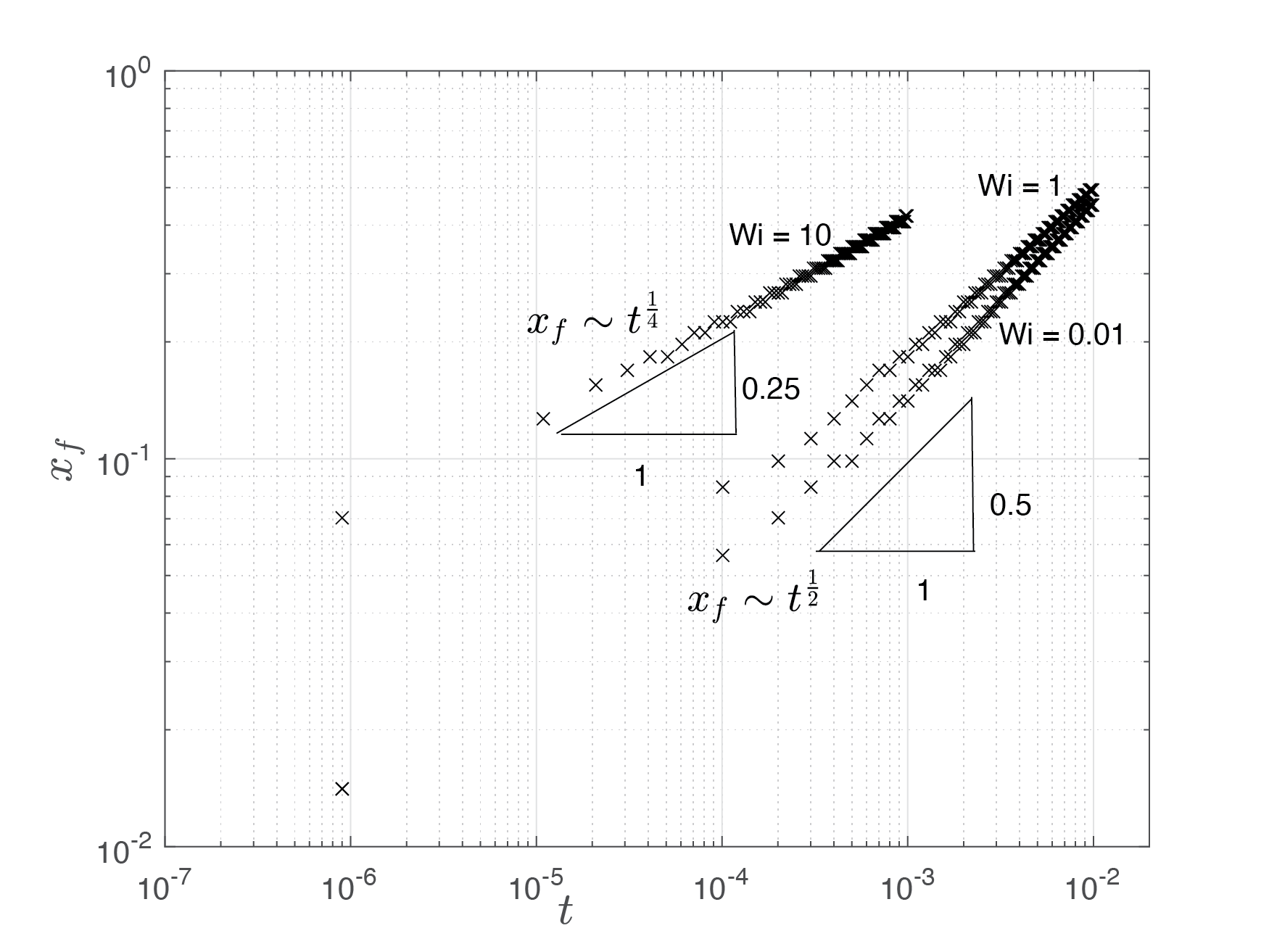}
\caption{Front width $x_f$ plotted as a function of $t$ for values of $Wi$. The value of $\beta$ is fixed at $0.4$. The definitions of the FSI parameter $\beta$, Weissenberg number $Wi$, and non-dimensional time $t$ are given by Eq. \eqref{defnP} and Eq.\eqref{defnDe} . }
\label{xfVstime}
\end{figure}

To be more quantitative, we define the front width $x_f$ as coordinate at which the pressure drops below $10^{-3}$.  In Fig.\ref{xfVstime}, we plot how $x_f$ varies in time for different $Wi$.  We see an interesting trend.  For low $Wi$ ($0 < Wi < 0.1$), the front appears to increase as $x_f \sim t^{1/2}$, while for high $Wi$ ($Wi \sim O(10)$) the front appears to increase as $x_f \sim t^{1/4}$. At first glance, this data seems at odds with the observation that viscoelasticity hastens the time to reach equilibrium.  However, we see from the plot  that the prefactor in front of the the power law is an order of magnitude larger for strongly viscoelastic fluids. We will obtain analytical estimates for such scaling relationships in the next section. 

\subsection{ Scaling and similarity solution for pressure controlled flow}\label{simsol}
Let us look at PDE Eq.\eqref{height_evolution} with the substitution $h = 1 + \beta p$:

\begin{subequations}
    \begin{align}
        &\beta \frac{\partial p}{\partial t} = \frac{\partial}{\partial x}\left( \frac{1}{3} h^{3}\frac{\partial p}{\partial x} + \frac{2\varepsilon Wi^{2}  }{5} h^{5}\left(\frac{\partial p}{\partial x}\right)^{3}\right); \qquad h = 1 + \beta p \label{height_evolution_substituted} \\
        &p(x, t=0) = 0; \quad p(x=0,t) = 1; \quad p(x=1,t) = 0 \label{height_evolution_substituted_BCs}
    \end{align}
\end{subequations}

Under specific conditions, the PDE admits a similarity solution where one can obtain scaling relationships for the front width ($x_f$) and the peeling time ($t_f$) -- i.e, the time it takes for the front to reach the conduit exit. Tables \ref{tbl:scaling_summary1} and \ref{tbl:scaling_summary2} summarize the scaling laws for different parameter regimes.  Below describes how one obtains such expressions.\\

\begin{table}
\centering
  \begin{tabular}{|p{0.1\textwidth}|p{0.1\textwidth}|p{0.1\textwidth}|}
  \hline
     & Newtonian regime ($\varepsilon Wi^2 \ll 1$) & non-Newtonian regime ($\varepsilon Wi^2 \gg 1$) \\
    \hline
    Non-dimensional peeling time ($t_f$) & $\beta $ & $\beta/(\varepsilon Wi^2)$ \\
    \hline
    Non-dimensional front width ($x_f$) & $ \sqrt{t/t_f} $ & $\left(t/t_f \right)^{1/4}$ \\
        \hline
     Dimensional peeling time ($t_f^*$) & $t_{FSI}^* $ & $t_{FSI}^*/(\varepsilon Wi^2)$ \\
    \hline
    Dimensional front width ($x_f^*$) & $ l^* \sqrt{ t^*/t_f^*}$ & $l^* \left(t^*/t_f^* \right)^{1/4}$ \\
    \hline
\end{tabular}%
  \caption{Scaling relationships for front width and peeling time for an sPTT fluid with moderate conduit deformations $\beta \sim O(1)$. The definitions of the FSI parameter $\beta$, Weissenberg number $Wi$, and non-dimensional time $t$ are given by Eq.\eqref{defnDe} and Eq. \eqref{eq:non_dim_time_sPTT}.  The  Newtonian FSI timescale in the small deformation limit is is $t_{FSI}^* = \frac{\eta^*}{\delta^2 h_0^* \kappa^*}$.}
  
  \vspace{-1em}
\label{tbl:scaling_summary1}
\end{table}

\begin{table}[h!]
\centering
  \begin{tabular}{|p{0.1\textwidth}|p{0.1\textwidth}|p{0.1\textwidth}|}
  \hline
     & Newtonian regime ($\varepsilon Wi^2\beta^2 \ll 1$) & non-Newtonian regime ($ \varepsilon Wi^2\beta^2 \gg 1$) \\
    \hline
    Non-dimensional peeling time ($t_f$) & $\beta^{-2} $ & $1/(\beta^4 \varepsilon Wi^2)$ \\
    \hline
    Non-dimensional front width ($x_f$) & $ \sqrt{t/t_f} $ & $\left(t/t_f \right)^{1/4}$ \\
        \hline
     Dimensional peeling time ($t_f^*$) & $t_{FSI}^*/\beta^3$ & $t_{FSI}^*/(\beta^5 \varepsilon Wi^2)$ \\
    \hline
    Dimensional front width ($x_f^*$) & $ l^* \sqrt{ t^*/t_f^*}$ & $l^* \left(t^*/t_f^* \right)^{1/4}$ \\
    \hline
\end{tabular}%
  \caption{Scaling relationships for front width and peeling time for strong conduit deformations $\beta \gg 1$.  The definitions of the FSI parameter $\beta$, Weissenberg number $Wi$, and non-dimensional time $t$ are given by Eq.\eqref{defnDe} and Eq. \eqref{eq:non_dim_time_sPTT}.  The  Newtonian FSI timescale in the small deformation limit is $t_{FSI}^* = \frac{\eta^*}{\delta^2 h_0^* \kappa^*}$}
  \vspace{-1em}
\label{tbl:scaling_summary2}
\end{table}

\subsubsection{Moderate conduit deformation ($\beta \sim O(1)$ or smaller)}\label{sec:moderate_sptt}

When the conduit's deformations are moderate ($\beta \sim O(1)$), the relative contribution of the two terms on the right hand side of Eq.\eqref{height_evolution_substituted} depend on the quantity $\varepsilon Wi^2$.   The conduit's FSI is dominated by Newtonian fluid rheology when $\varepsilon Wi^2 \ll 1$, while the FSI dominated by non-Newtonian rheology when $\varepsilon Wi^2 \gg 1$.  Below describes similarity solutions under these regimes.\\

\underline{Newtonian FSI}:  Let us inspect the case of Newtonian FSI -- i.e., $\varepsilon Wi^2 \ll 1$.  Upon taking this limit, the PDE Eq.\eqref{height_evolution_substituted} becomes:

\begin{equation}
\label{height_evolution_Newtonian}
\beta \frac{\partial p}{\partial t} = \frac{\partial}{\partial x}\left( \frac{1}{3} h^{3}\frac{\partial p}{\partial x}\right);  \qquad h = 1 + \beta p \\
\end{equation}
with the same boundary conditions as before (Eq.\eqref{height_evolution_substituted_BCs}).  For early times $t \ll \beta$, the height profile admits a similarity solution.  The profile looks like a front that propagates from the conduit's inlet to the exit, with the front's width scaling as $x_f \sim \sqrt{t/\beta}$.  The front reaches the end when $x_f \sim O(1)$, which gives an expression for the peeling time as $t_f \sim \beta$.  Using these ideas, we write the form of the solution as follows:


\begin{figure}

\centering
  \includegraphics[width=0.5\textwidth]{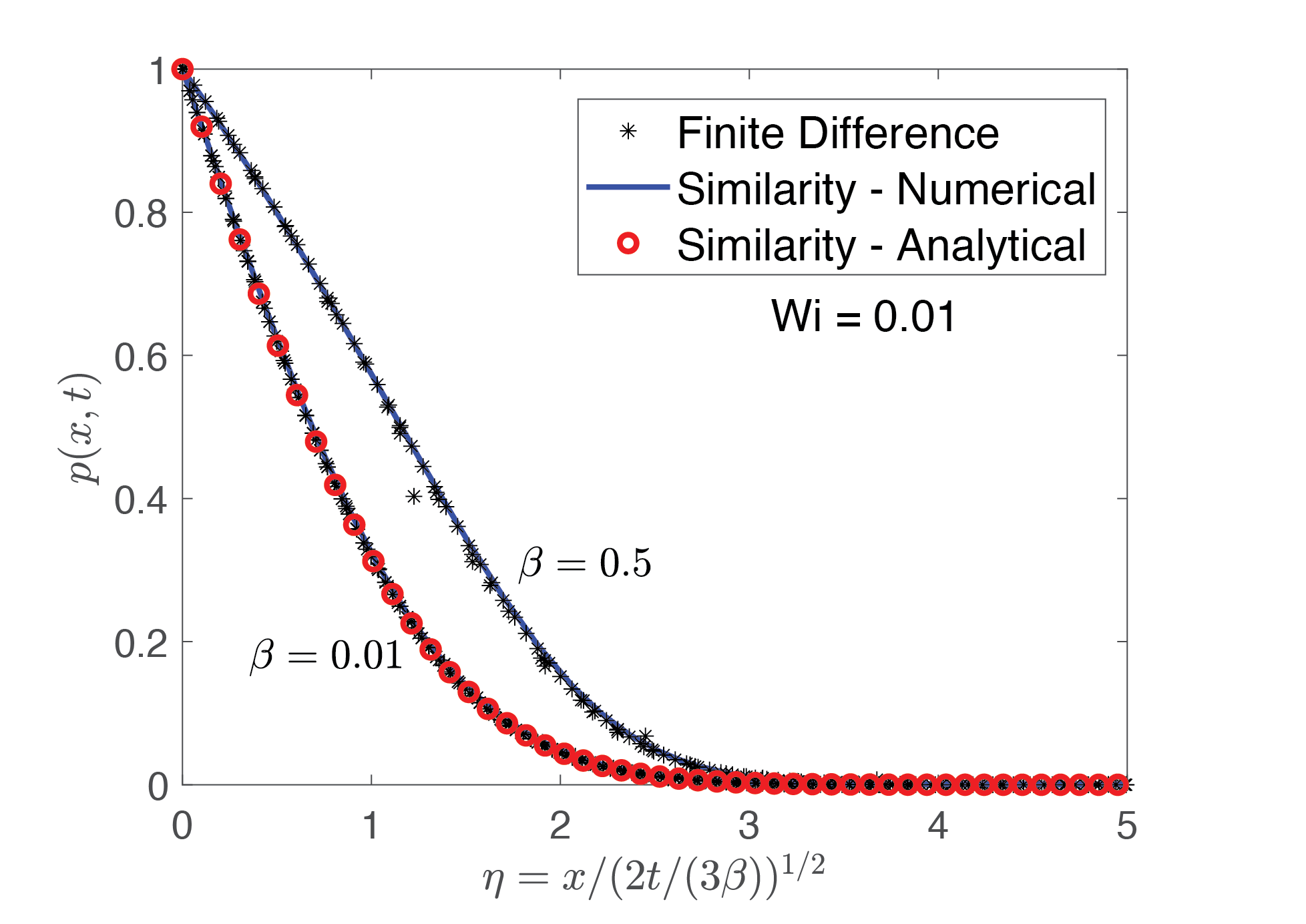}

\caption{  Similarity solution for Newtonian FSI ($\varepsilon Wi^2 \ll 1$) at moderate sheet deformations ($\beta \sim O(1))$ or smaller).  The similarity solution is plotted against the numerical finite difference solution at different time instances.  For $\beta = 0.01$, the time instances plotted are $t = 5 \times 10^{-5},5\times 10^{-4},$ and $10^{-3}$.  For $\beta = 0.5$, the time instances are $t = 5 \times 10^{-5},5\times 10^{-4}, 10^{-3}$, and $2 \times 10^{-2}$.  For both plots, $Wi = 0.01$ and $\varepsilon = 0.01$.  For $\beta = 0.01$, the leading order analytical solution for $\beta \ll 1$ is also plotted ($p = \text{erfc}(\eta/\sqrt{2})$). The definitions of the FSI parameter $\beta$, Weissenberg number $Wi$, and non-dimensional time $t$ are given by Eq.\eqref{defnDe} and Eq. \eqref{eq:non_dim_time_sPTT}. } 
\label{fig:similarity_lowkDe}
\end{figure}

\begin{equation}
    p = f(\eta);   \qquad \eta \equiv \frac{x\sqrt{3\beta}}{\sqrt{2t}}
\end{equation}

The above transformation converts the PDE Eq.\eqref{height_evolution_Newtonian} into an ODE:
\begin{subequations}
\begin{align}
    &\frac{d}{d \eta}\left( h^3 \frac{df}{d\eta}\right) + \eta \frac{df}{d\eta} = 0;   \qquad h = 1 + \beta f \\
    &f(0) = 1; \quad f(\infty) = 0    
\end{align}
\end{subequations}

This ODE has an analytical solution for $\beta \ll 1$:  $f = \text{erfc}(\eta/\sqrt{2})$.  For $\beta \sim O(1)$, one can numerically solve the ODE using a shooting method, where one applies the boundary condition at infinity at a sufficiently large value of $\eta$ (typically $\eta > 10$). We increase $\eta$ up to a point where the results become insensitive to the value of $\eta$. At this value, the similarity plot matches the numerical solution fairly well. Fig.\ref{fig:similarity_lowkDe} plots the similarity solution versus the fully resolved finite difference solution from the previous section.  For the conditions stated -- i.e., Newtonian fluid ($\varepsilon Wi^2 \ll 1$) and early times ($t \ll \beta$), the similarity solution captures the deformation profile well.\\

\underline{Non-Newtonian FSI}:    When the FSI is dominated by the non-Newtonian rheology, $\varepsilon Wi^2 \gg 1$.  In this limit, the PDE Eq.\eqref{height_evolution_substituted} for the pressure becomes:

\begin{equation}
\label{height_evolution_Newtonian}
\beta \frac{\partial p}{\partial t} = \frac{\partial}{\partial x}\left( \frac{2\varepsilon Wi^{2}}{5} h^{5}\left(\frac{\partial p}{\partial x}\right)^{3}\right); \qquad h = 1 + \beta p \\
\end{equation}
with the same boundary conditions as before (Eq.\eqref{height_evolution_substituted_BCs}).  For early times $t \ll \beta/(\varepsilon Wi^2)$, the height profile admits a similarity solution.  The profile propagates as a front with thickness scaling as $x_f \sim \left( \varepsilon Wi^2 t/ \beta \right)^{1/4}$.  The front reaches the end when $x_f \sim O(1)$, which gives the peeling time as $t_f \sim \beta/(\varepsilon Wi^2)$.  Using these ideas, we write the form of the solution:


\begin{equation}
    p = f(\eta);   \qquad \eta \equiv \frac{x (5\beta)^{1/4}}{(8 t \varepsilon Wi^2)^{1/4}}
\end{equation}

The above transformation converts the PDE \eqref{height_evolution_Newtonian} into an ODE:

\begin{subequations}
\begin{align}
    &\frac{d}{d \eta}\left( h^5 \left(\frac{df}{d\eta}\right)^3 \right) + \eta \frac{df}{d\eta} = 0;   \qquad h = 1 + \beta f \\
    &f(0) = 1; \quad f(\infty) = 0    
\end{align}
\end{subequations}

This ODE is different than the Newtonian ODE in that it is not well defined for the entire domain $\eta \in [0, \infty)$.  At a particular point $\eta_{max}$, the pressure gradient will switch sign:

\begin{equation*}
    \frac{df}{d\eta}(\eta_{max}) = 0 
\end{equation*}

Clearly, this behavior is aphysical so the similarity ODE can only be defined in the region $\eta \in [0, \eta_{max}]$.  What is physically happening is that around $\eta = \eta_{max}$, the pressure gradients are so small that the Newtonian contributions to the FSI start to matter.  Because the pressure is so small in this region and the region $\eta > \eta_{\max}$, the details are unimportant for all practical purposes.  Thus, one can reformulate the above ODE as follows:

\begin{figure}
\centering
\subfloat[]{
  \includegraphics[width=0.4\textwidth]{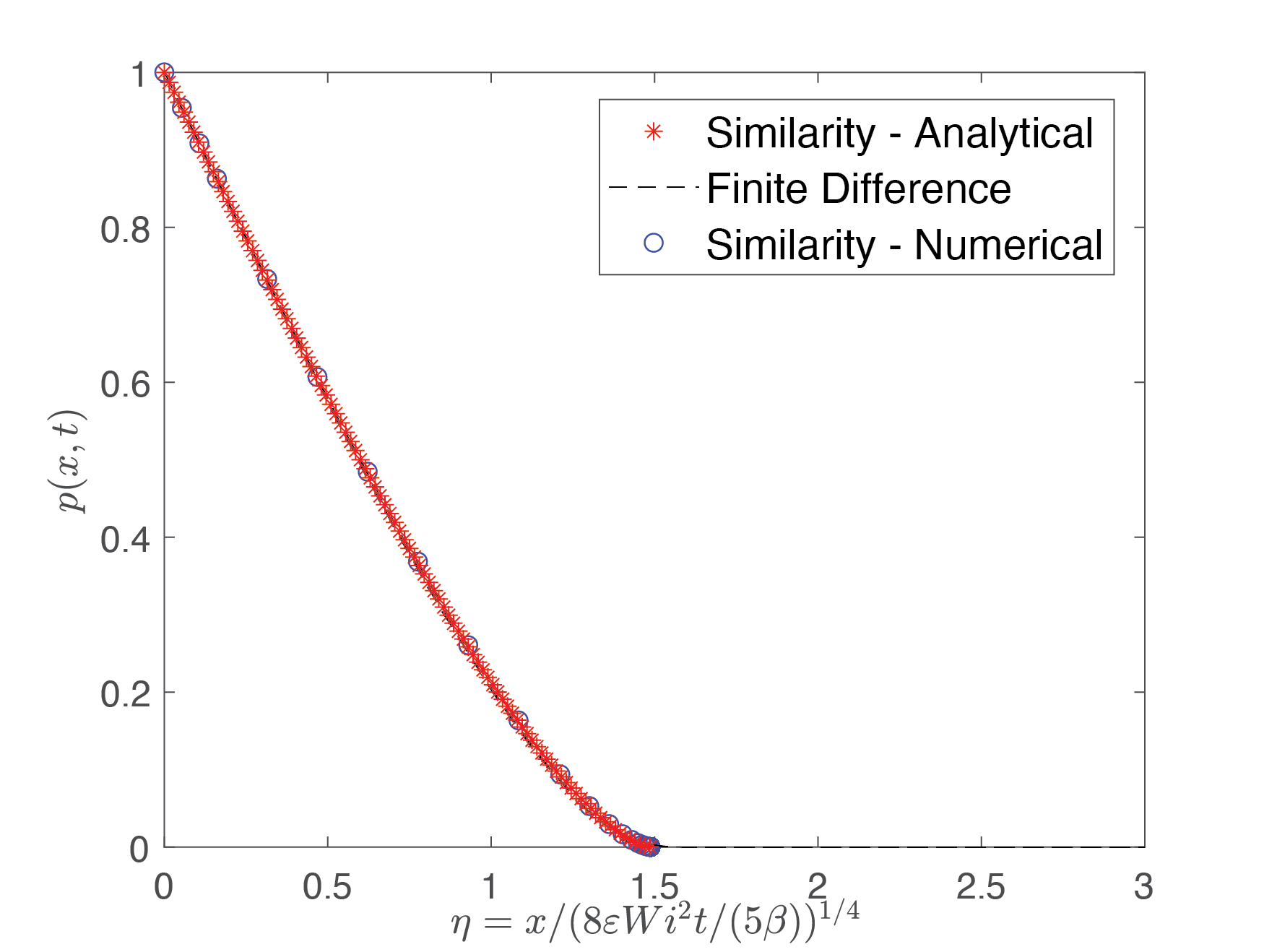}
}
\vspace{-1em}
\subfloat[]{
  \includegraphics[width=0.4\textwidth]{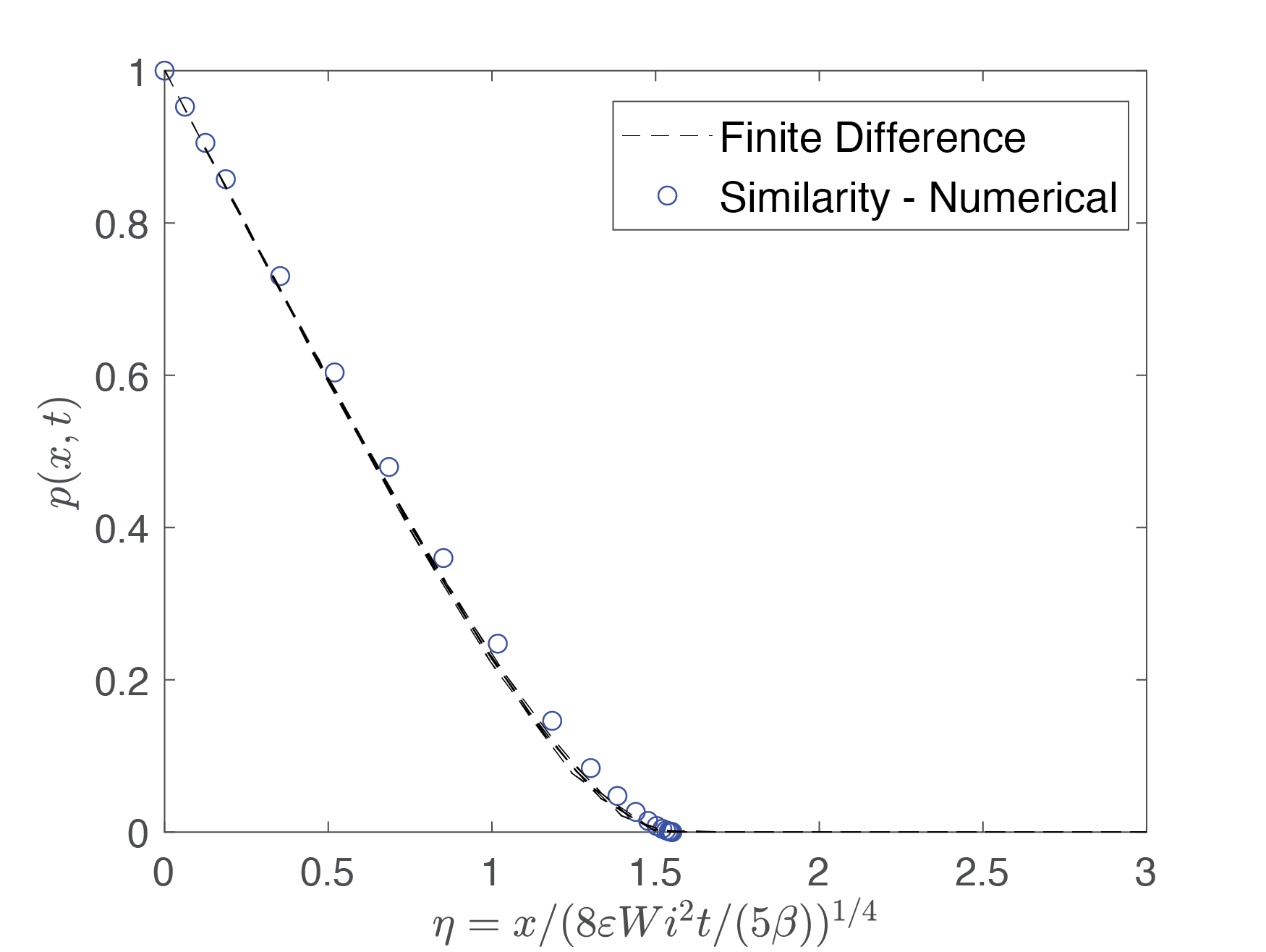}
}
\vspace{-1em}
\subfloat[]{
  \includegraphics[width=0.4\textwidth]{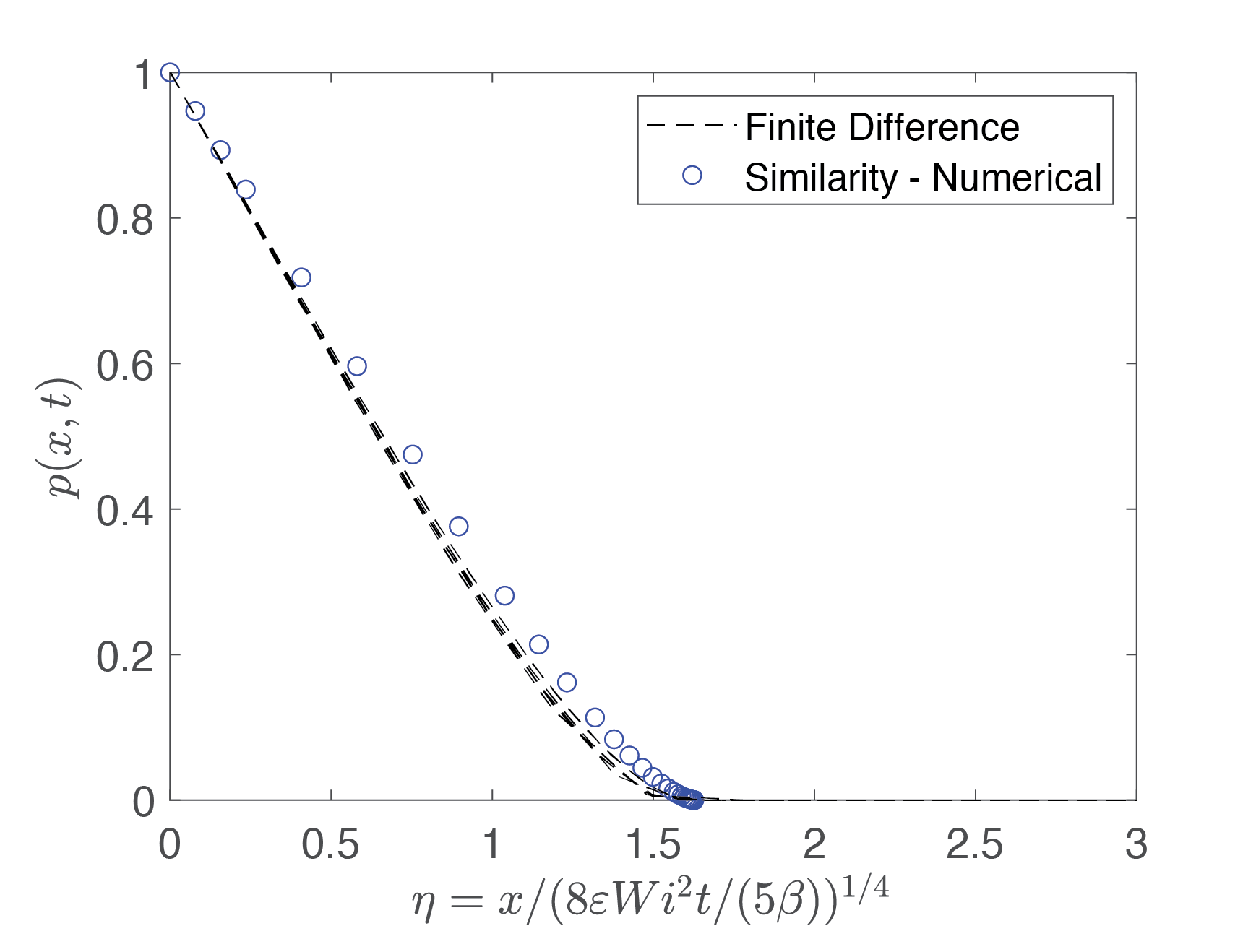}
}

\caption{Similarity solution for viscoelastic FSI ($\varepsilon Wi^2 \gg 1$) at moderate sheet deformations ($\beta \sim O(1))$ or smaller). (a) $\beta = 0.005$.  The similarity solution is plotted against the numerical finite difference solution at different time instances $t = 5 \times 10^{-7},10^{-6}$, and $10^{-5}$.  The red curve is the leading order analytical solution for $\beta \ll 1$ (Eq. \eqref{analytical_lowbeta_HighDe}).  (b) $\beta = 0.1$.  The similarity solution is plotted against the numerical solution at times $t = 1 \times 10^{-6},5 \times 10^{-6}$, and $10^{-4}$. (c) $\beta = 0.2$. The similarity solution is plotted against the numerical solution at times $t = 1 \times 10^{-6},5 \times 10^{-6}$, $10^{-4}$, and $10^{-3}$.  For all plots, $Wi = 20$ and  $\varepsilon = 0.01$. The definitions of the FSI parameter $\beta$, Weissenberg number $Wi$, and non-dimensional time $t$ are given by Eq.\eqref{defnDe} and Eq. \eqref{eq:non_dim_time_sPTT} }
\label{SimilaritysPTT_plots}
\end{figure}

\begin{subequations}
\begin{align}
    &\frac{d}{d \eta}\left( h^5 \left(\frac{df}{d\eta}\right)^3 \right) + \eta \frac{df}{d\eta} = 0;   \qquad h = 1 + \beta f \\
    &f(0) = 1; \quad f'(\eta_{max}) = 0; \qquad f(\eta_{max}) = 0    
\end{align}
\end{subequations}
where we note that the solution will have a small error in a thin region around the endpoint.  For the case of small deformations ($\beta \ll 1$), the above ODE has an analytical solution:

\begin{equation}\label{analytical_lowbeta_HighDe}
    f = 1 - \frac{\eta_{max}^2}{\sqrt{3}} \int_0^{\eta/\eta_{max}} \sqrt{1 - z^2} dz \qquad \eta_{max} = \left(\frac{4\sqrt{3}}{\pi}\right)^{1/2} \qquad \beta \ll 1
\end{equation}

For values of $\beta \sim O(1)$, one will have to numerically solve the above ODE.  The way we approach this task is to perform a nested shooting method.  We first define a function that takes an input $\eta_{max}$ and uses a shooting method to compute the solution to the above ODE with boundary conditions $f(0) = 1$ and $f'(\eta_{max}) = 0$.  The function outputs the value at the endpoint $f(\eta_{max})$.  One performs root finding on the output of this function until one obtains the value of $\eta_{max}$ at which $f(\eta_{max}) = 0$. Fig.\ref{SimilaritysPTT_plots} shows the solution to the similarity ODE and compares it to the fully resolved finite difference solution described previously. This figure encapsulates the similarity scaling discussed in Table \ref{tbl:scaling_summary1}. 

\begin{figure}
\centering
\subfloat{
  \includegraphics[width=0.5\textwidth]{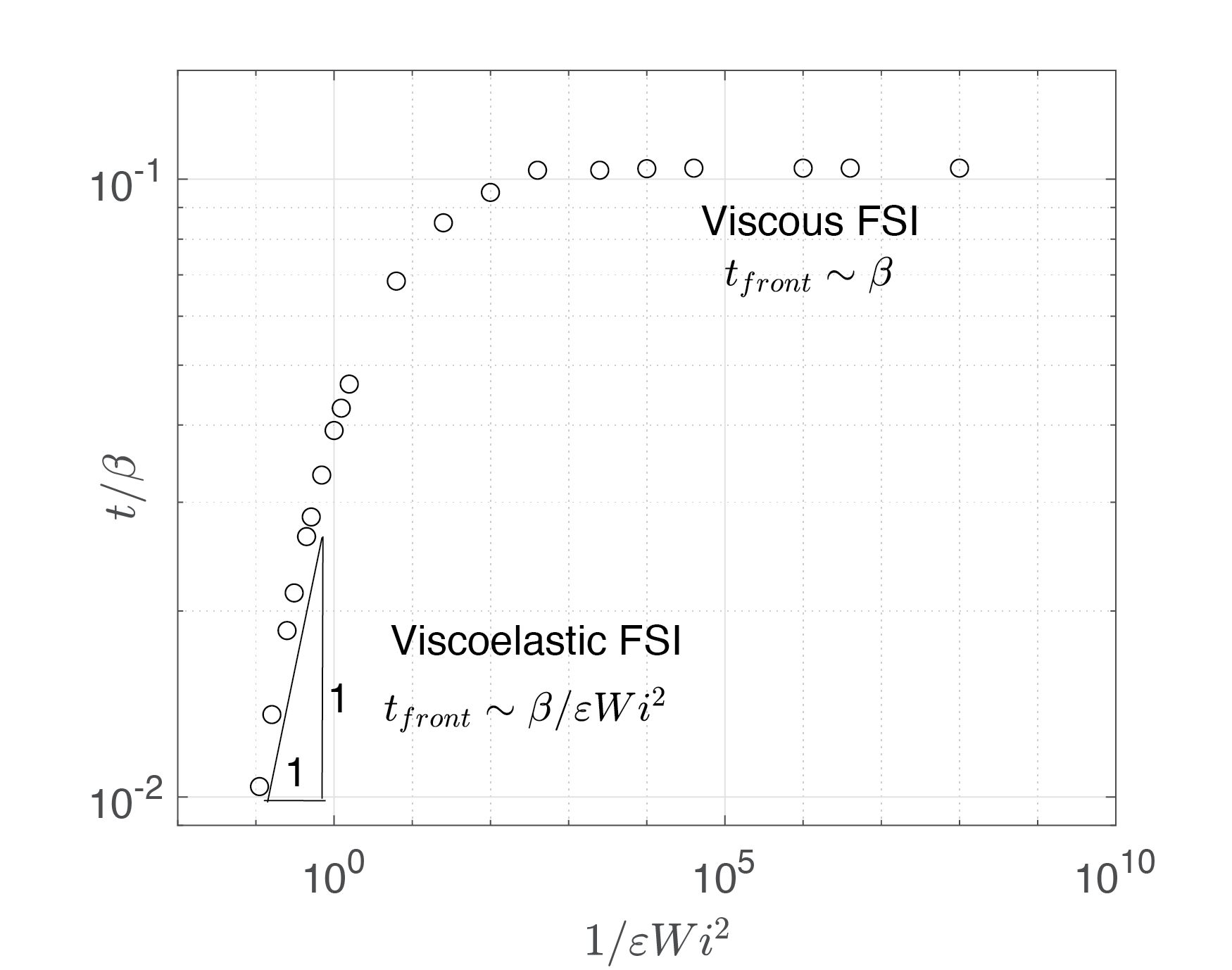}
}
\caption{ Summary of peeling time for moderate conduit deformations $(\beta \sim O(1)$ or smaller).  The front width ($x_{f}$) is the coordinate at which the scaled pressure $p$ drops to under $10^{-3}$.  The peeling time is defined as the time at which $x_{f} = 1$.The definitions of the FSI parameter $\beta$, Weissenberg number $Wi$, and non-dimensional time $t$ are given by Eq.\eqref{defnDe} and Eq. \eqref{eq:non_dim_time_sPTT} }
\label{regimeplot}
\end{figure}

 Fig.\ref{regimeplot} summarizes the scaling behavior observed in the numerical simulations for the peeling time at different Weissenberg numbers at moderate conduit deformations ($\beta \sim O(1)$ or smaller).  The figure clearly demarcates the two regimes of a strongly Newtonian fluid (where $t_f \sim \beta$) and a strongly viscoelastic fluid (where $t_f \sim \beta/(\varepsilon Wi^2)$).  At moderate viscoelasticity ($\varepsilon Wi^2 \sim O(0.1)$), neither scaling regime holds and one must resort to a full numerical solution.  The next section will discuss the regime of strong conduit deformations ($\beta \gg 1$). 


\subsubsection{Strong conduit deformations ($\beta \gg 1$)}

When the conduit's deformation is strong, the conduit height scales as $h \sim \beta \gg 1$ rather than $h \sim O(1)$ as stated in the previous section.  This fact alters the relative magnitude of the Newtonian and viscoelastic terms in the pressure PDE eq (\ref{height_evolution_substituted}).  Newtonian FSI now occurs when $\varepsilon Wi^2 \beta^2 \ll 1$, while viscoelsatic FSI occurs in the opposite limit $\varepsilon Wi^2\beta^2 \gg 1$.  Below is a brief summary of the similarity solutions in both limits.\\

\underline{Newtonian FSI}:  Under the Newtonian limit  $\varepsilon Wi^2\beta^2 \ll 1$, the limiting PDE for the pressure profile is the same as eqn (\ref{height_evolution_Newtonian}).  However, since the conduit's height scales as $h \sim \beta$, the peeling front width and the peeling time are different.  The front's width scales as $x_f \sim \sqrt{t\beta^2}$.  The peeling time is determined when $x_f \sim O(1)$, which occurs at $t_f \sim \beta^{-2}$.  

\begin{figure}
\centering
\subfloat{
  \includegraphics[width=0.5\textwidth]{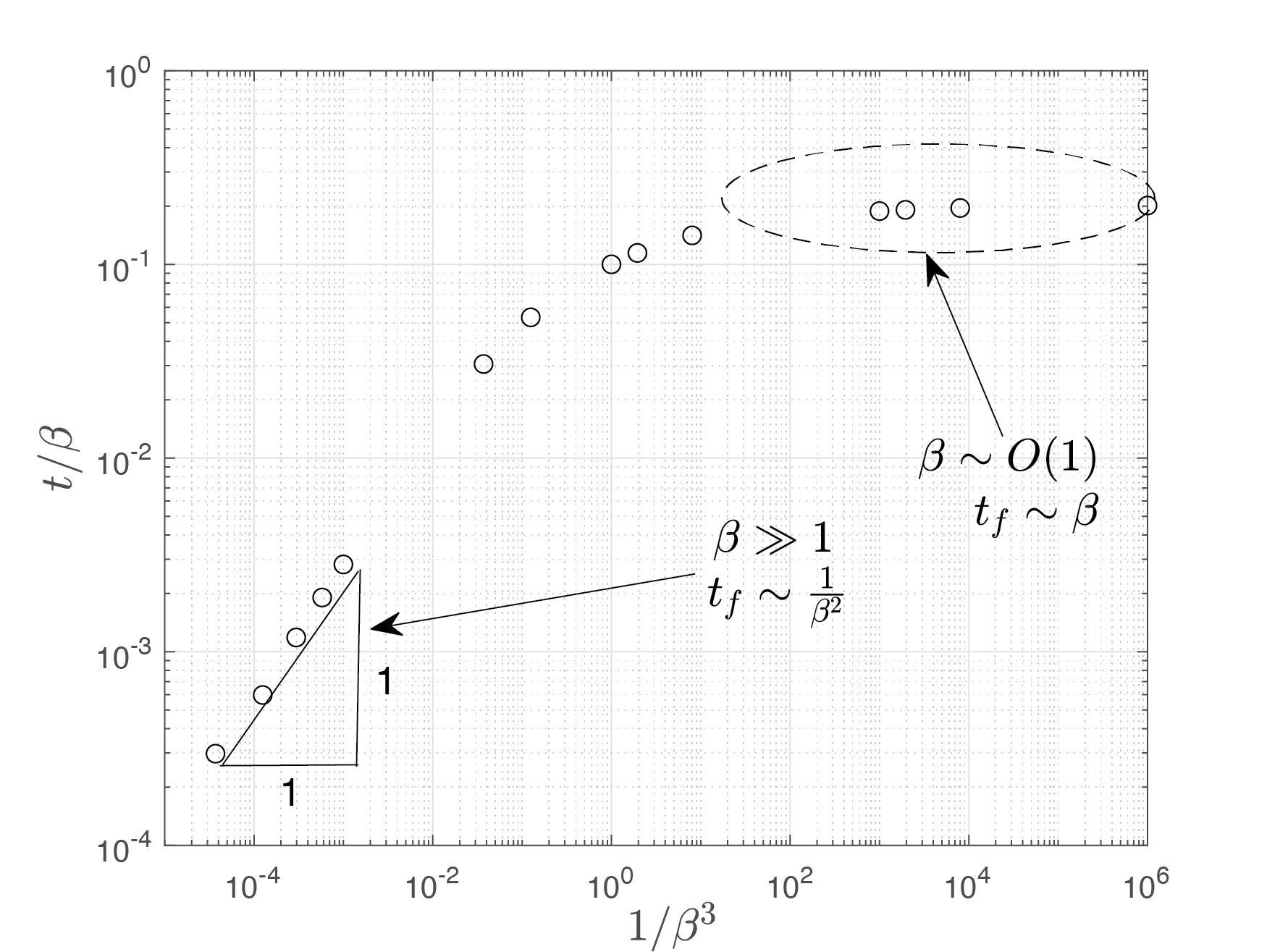}
}
\caption{Peeling time versus FSI parameter $\beta$ for the Newtonian FSI regime ($\varepsilon Wi^2 \ll 1$ for $\beta \sim O(1)$ or smaller, and $\varepsilon Wi^2\beta^2 \ll 1$ for $\beta \gg 1$). In the plot, we fix $Wi = 1$ and $\varepsilon = 0.01$ and vary $\beta$.  The definition of $\beta$ for this problem is given in Eq.\eqref{defnDe}.}
\label{timedependenceBeta}
\end{figure}

We note that the peeling time in this strong deformation limit is markedly different than the moderate deformation case discussed previously, where $t_f \sim \beta$. Fig.\ref{timedependenceBeta} shows numerical simulations of the peeling time for different values of $\beta$ that clearly demarcates the two regimes.  The non-monotonic trend in peeling time on the conduit deformation $\beta$ is a non-intuitive finding that arises due to the competition between lubrication pressure ($p_{lub} \sim \frac{1}{h^3} \frac{\partial h}{\partial t}$) and the sheet’s elastic stresses ($p_{elas} \sim (h-1) \beta$).  At large conduit deformations, the lubrication pressure is greatly attenuated, which gives rise to the trends above.

To obtain a similarity solution in the regime of strong conduit deformation $(\beta \gg 1)$, we make the following transformation:

\begin{equation}
    p = f(\eta);   \qquad \eta \equiv \frac{x\sqrt{3}}{\sqrt{2t\beta^2}}
\end{equation}

which converts the PDE Eq.\eqref{height_evolution_Newtonian} into the ODE:
\begin{subequations}
\begin{align}
    &\frac{d}{d \eta}\left( \tilde{h}^3 \frac{df}{d\eta}\right) + \eta \frac{df}{d\eta} = 0;   \qquad \tilde{h} = f + \beta^{-1} \\
    &f(0) = 1; \quad f(\infty) = 0    
\end{align}
\end{subequations}
This equation is not well-defined on the entire domain $\eta \in [0, \infty)$.  There is a specific value of $\eta$ at which the pressure drops to zero:

\begin{equation*}
    f(\eta_{max}) = 0
\end{equation*}

We thus replace the boundary condition at infinity with the one above.  The value of $\eta_{max}$ is obtained from simulation.

\begin{figure}

\centering
\includegraphics[width=0.5\textwidth]{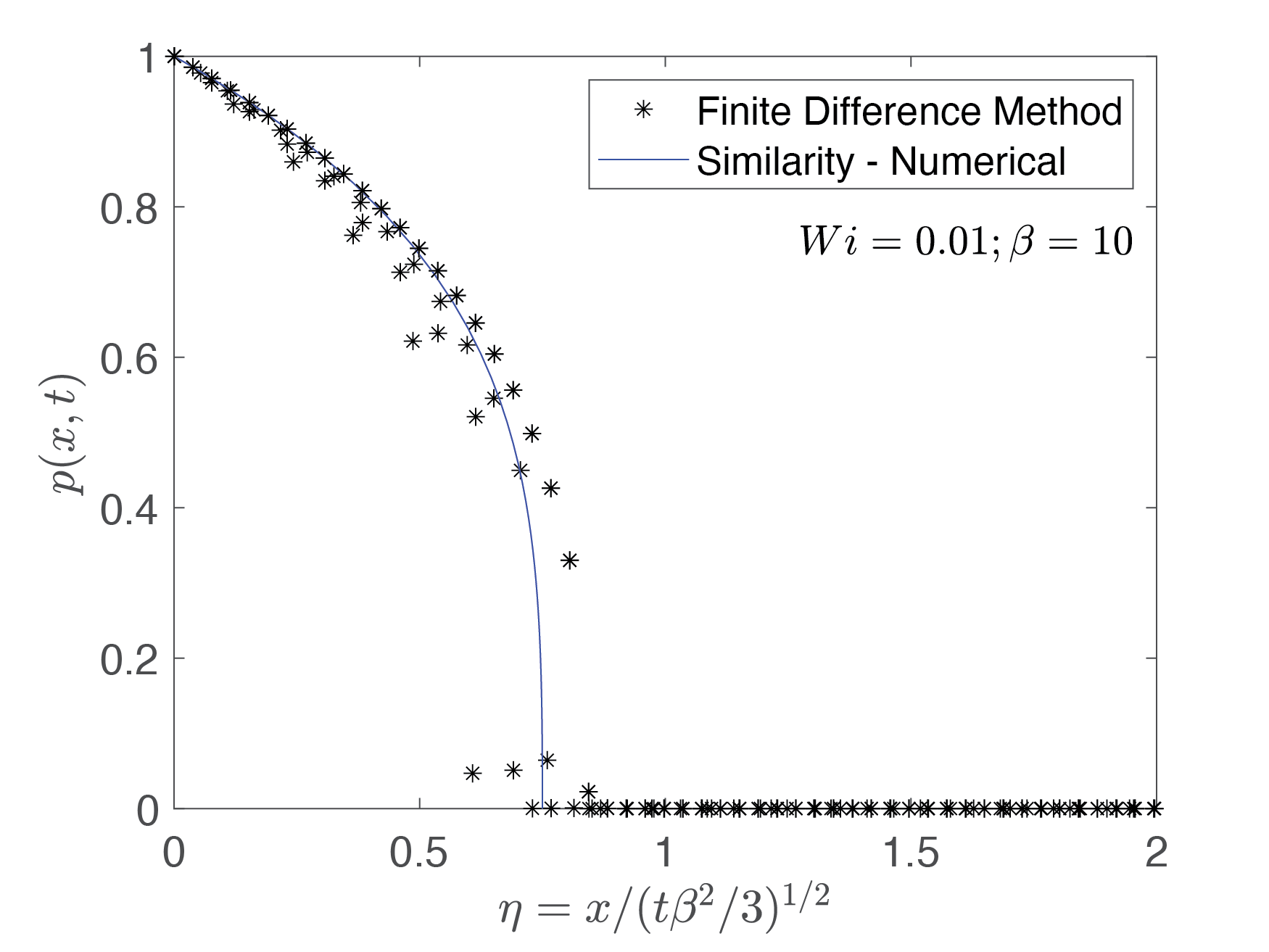}
\caption{Similarity solution for Newtonian FSI ($\varepsilon Wi^2 \beta^2 \ll 1$) for strong sheet deformations ($\beta \gg 1)$).  The similarity solution is plotted against the numerical finite difference solution at different time instances $t = 10^{-5}, 10^{-4},10^{-3}$. The definitions of the FSI parameter $\beta$, Weissenberg number $Wi$, and non-dimensional time $t$ are given by Eq.\eqref{defnDe} and Eq. \eqref{eq:non_dim_time_sPTT}. }
\label{fig:similarity_strong_def_Newtonian}
\end{figure}

Fig.\ref{fig:similarity_strong_def_Newtonian} plots the similarity solution in the high $\beta$ limit compared to the finite difference simulations.  Overall, we see a reasonable collapse of the numerical data to the similarity curve, although there is some scatter.  We also note that the similarity solution is qualitatively different than the moderate deformation case ($\beta \sim O(1)$ in Fig.\ref{SimilaritysPTT_plots}).  Here, the shape of the front is concave down, whereas the shape for the moderate deformation case is concave up.\\

\underline{Non-Newtonian FSI}:  Table \ref{tbl:scaling_summary1} lists the scaling relationships for non-Newtonian FSI in the regime of strong conduit deformation ($\beta \gg 1$).  Although one can develop a similarity solution in this regime, we believe that this situation might not be realizable using the current model.  At high conduit deformations, the value of $\delta Wi$ will likely no longer be small (likely $O(0.1)$ or larger), making the upper convected derivative term in Eq. \eqref{eq:sPTT} important.  In this situation, the relaxation of the polymer chain will be comparable to the average transit time in the conduit, giving a non-quasi steady flow profile \cite{boyko_stone_2022}. We will not discuss this situation in this manuscript.


\section{Lubrication analysis - Power law fluid}\label{lubrication_analysis:powerlaw}
\subsection{Lubrication Equations}
The setup is exactly the same as described in section \ref{setup} except that the non-Newtonian fluid is now a power law fluid with power law index $n$ and consistency index $K^*$.  For an initial gap thickness $h_0^*$ and average fluid velocity $U^*$, the effective viscosity scale is:

\begin{equation}\label{powerlawViscosity}
    \eta_{eff}^* = K^* \left( \frac{U^*}{h_0^*} \right)^{n-1}
\end{equation}
When performing the lubrication analysis discussed before, we will keep the same definitions for the non-dimensional lengths, velocities, time, pressure, and stresses as in Eq.\eqref{eq:dist_nondimensional}-Eq.\eqref{eq:stress_nondimensional}.  The relationship between the lubrication pressure scale $P^*_{lub}$ and the velocity scale $U^*$ will have a similar form as Eq.\eqref{eq:pressure_scale}, except that one replaces the viscosity of the fluid by the effective viscosity $\eta^*_{eff}$:
 
 \begin{equation}\label{powerlawPressure}
     P^*_{lub} =  \frac{\eta_{eff}^*U^*}{\delta h_0^*}
 \end{equation}
 Depending on the problem at hand (pressure controlled or flow controlled), one of the two scales $P^*_{lub}$ or $U^*$ will be specified with the other determined by the above two relationships (Eq. \eqref{powerlawViscosity}, Eq.\eqref{powerlawPressure}).  The lubrication approximation will hold if the conduit is long and slender ($\delta \ll 1$) and the effective Reynolds number $Re_{eff} \ll \delta^{-1}$.  Here, the effective Reynolds number is the Reynolds number using the effective viscosity defined above:  $Re_{eff} = \rho^* U^* h_0^*/\eta_{eff}^*$.

In non-dimensional form, the lubrication equations for the momentum balance will take the form:

\begin{equation}
    \frac{\partial \tau_{xy}}{\partial y} = \frac{\partial p}{\partial x};   \qquad \frac{\partial p}{\partial y} = 0
\end{equation}
where
\begin{equation} \label{eq;power_law_shear_stress}
    \tau_{xy} = \left|\frac{\partial u}{\partial y}\right|^{n-1} \frac{\partial u}{\partial y}
\end{equation}

We substitute the above equation Eq. \eqref{eq;power_law_shear_stress} into the first equation of the momentum balance and integrate to find the velocity field $u$.  In doing this procedure, we apply the boundary conditions that there is no shear stress at the center of the conduit ($y=0$) and no velocity at the top ($y=h$).  This procedure yields:

\begin{equation}
    u = \frac{n}{1+n} \left(- \frac{\partial p}{\partial x} \right)^{1/n} \left( h^{\frac{1+n}{n}} - \left| y\right|^{\frac{1+n}{n}}  \right)     
\end{equation}
To derive the equation for the height of the conduit, we examine the continuity equation $\frac{\partial u}{\partial x} + \frac{\partial v}{\partial y} = 0$ and integrate it over the height of the conduit from $y = 0$ to $y = h$.  Noting that $v = \frac{\partial h}{\partial t}$ at the top wall, we obtain:

\begin{equation} \label{eq:PDE_powerlaw}
    \frac{\partial h}{\partial t} = -\frac{n}{2n+1} \frac{\partial}{\partial x} \left(  h^{\frac{2n+1}{n} } \left(-\frac{\partial p}{\partial x} \right)^{1/n}     \right)
\end{equation}

This equation is the PDE that describes the height distribution in the conduit.  It is coupled with the deformation relationship 

\begin{equation}
    h = 1 + \beta p
\end{equation}
where $\beta$ is the channel's FSI parameter. 

Eq.\eqref{eq:PDE_powerlaw} governs the evolution of the elastic deformation of the sheet. The flow rate controlled case at steady state has been discussed in detail by Anand et.al.\cite{anand2019non}. We will proceed to solve the transient problem for the pressure controlled dynamics. 
\subsection{Results -- Pressure controlled case}
\subsubsection{Definitions}
Similar to Section \ref{pressurecontrolled}, we will examine a problem where a stagnant conduit at zero pressure is subject to a pressure impulse at its inlet at time $t^* = 0$.  The inlet of the conduit is clamped at pressure $\Delta P^*$ while the outlet is clamped at zero pressure.  If we specify $\Delta P^*$, this will set the pressure scale, velocity scale, and effective viscosity scale for the non-dimensional variables in eqns Eq. \eqref{eq:dist_nondimensional}-Eq.\eqref{eq:vel_nondimensional}:
\begin{equation}\label{powerlawnondimensionalize}
    P^*_{lub} = \Delta P^*; \qquad U^* = h_0^* \left( \frac{\Delta P^* \delta}{K^*} \right)^{1/n} \qquad \eta_{eff}^* = K^* \left( \frac{\Delta P^* {\delta}}{K^*} \right)^{\frac{n-1}{n}}
\end{equation}
The FSI parameter for this problem is:
\begin{equation}\label{powerlawbeta}
   \beta = \frac{\Delta P^*}{h_0^* \kappa^*}
\end{equation}
and the dimensionless time is:
\begin{equation}\label{powerlawtime}
 t = t^* \frac{U^*}{l^*} = {\delta} t^* \left( \frac{\Delta P^* {\delta}}{K^*} \right)^{1/n}
\end{equation}
In non-dimensional form, the PDE that describes the height evolution of the channel is given by Eq.\eqref{eq:PDE_powerlaw} subject to the condition that $h = 1 + \beta p$.  The boundary conditions are that the pressure is $p(x=0,t) = 1$ and $p(x=1,t) = 0$, with the initial condition that $p(x,t=0) = 0$.





\subsubsection{Numerical results}
In this section, we solve the PDE given by Eq.\eqref{eq:PDE_powerlaw}. Our numerical procedure is the same as discussed in Section \ref{numsol} except that we use an \st{strongly} implicit Crank-Nicolson time-stepping scheme which is thoroughly described in reference \cite{Ghodgaonkar2019}. 

\begin{figure} 
\centering
\subfloat{
  \includegraphics[width=0.5\textwidth]{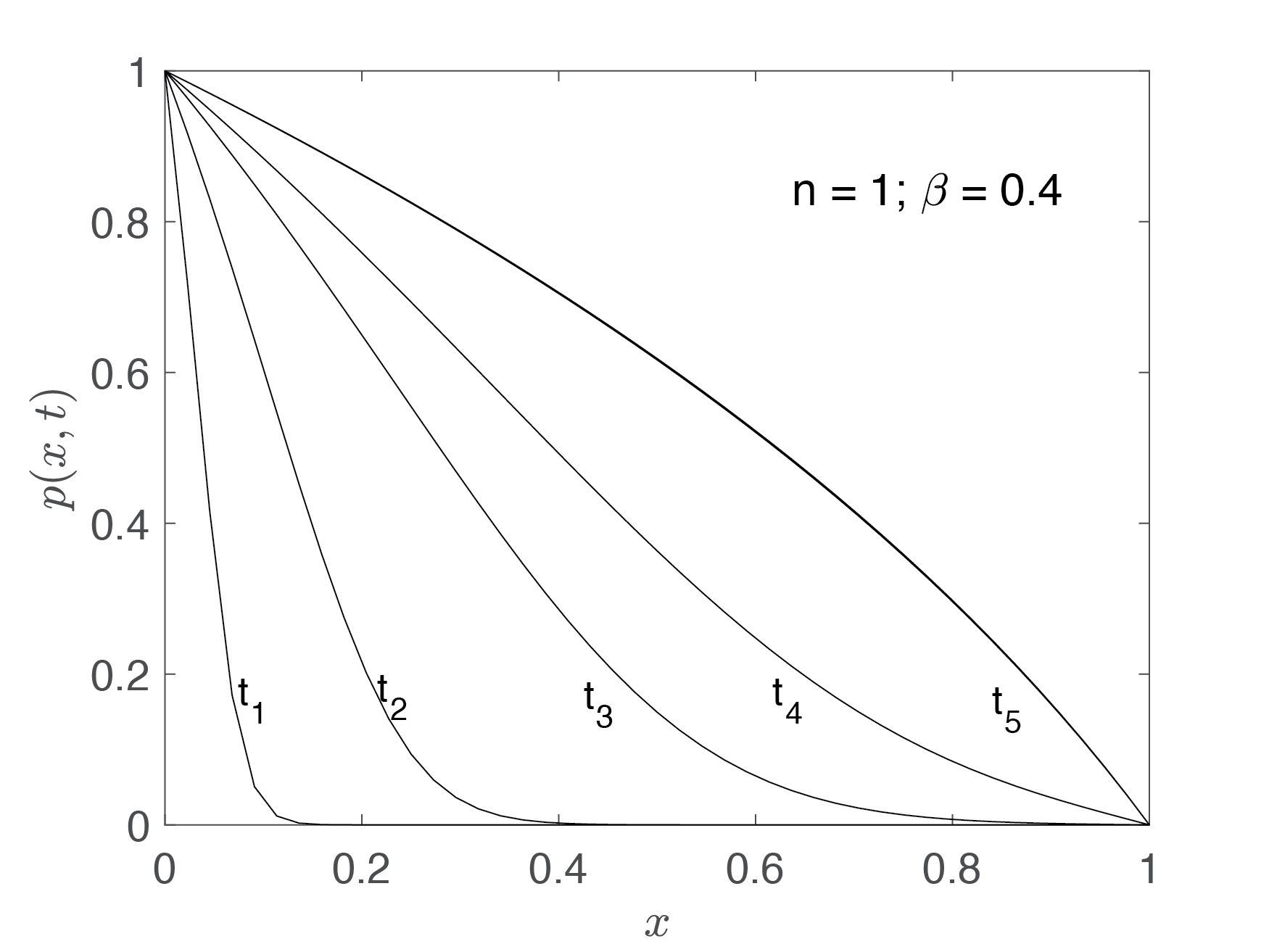}
}
\vspace{1em}
\subfloat{
  \includegraphics[width=0.5\textwidth]{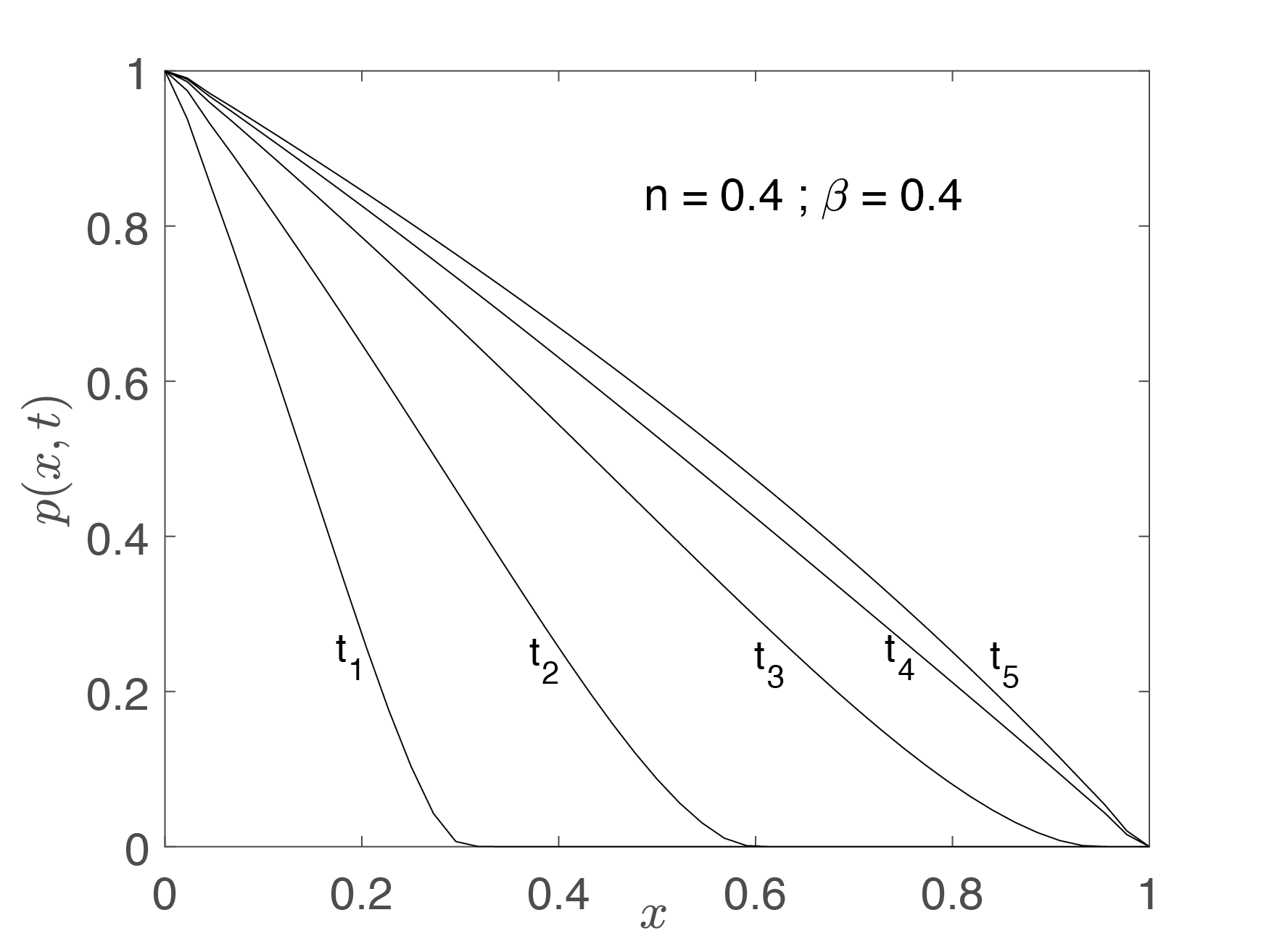}
}

\caption{Numerical solution of transient pressure profile for (a) Newtonian fluid ($n = 1$) and (b) moderately shear thinning fluid ($n = 0.4$).  The time instances sampled correspond to $t_{1} = 0.001\beta$, $t_{2} = 0.01\beta$, $t_{3} = 0.05\beta$,  $t_{4} = 0.1\beta$, and $t_{5} = 2.5\beta$. The FSI parameter is $\beta =0.4$.  Note -- the definition of FSI parameter $\beta$ and non-dimensional time $t$ are given by eqns (\ref{powerlawbeta}-\ref{powerlawtime}).}
\label{fig:shearthinning}
\end{figure}

Fig.\ref{fig:shearthinning} plots the pressure profile at different time instances for a Newtonian fluid and a moderately shear thinning fluid ($n = 0.4$).  The results indicate that shear thinning causes the front to propagate more quickly. The next section goes through scaling analyses and similarity solutions for this problem.

\subsubsection{Similarity solution}\label{simsolPowerLaw}

Let us look at PDE Eq. \eqref{eq:PDE_powerlaw} with the substitution $h = 1 + \beta p$:

\begin{subequations}
    \begin{align}
        & \beta \frac{\partial p}{\partial t} = -\frac{n}{2n+1} \frac{\partial}{\partial x} \left(  h^{\frac{2n+1}{n} } \left(-\frac{\partial p}{\partial x} \right)^{1/n}     \right); \qquad h = 1 + \beta p \label{height_PL_evolution_substituted} \\
        &p(x, t=0) = 0; \quad p(x=0,t) = 1; \quad p(x=1,t) = 0 \label{height_PL_evolution_substituted_BCs}
    \end{align}
\end{subequations}

We will follow a similar procedure as Section \ref{simsol} to determine a similarity solution for moderate conduit deformations ($\beta \sim O(1)$ or smaller) and large conduit deformations ($\beta \gg 1)$.  Table \ref{tbl:scaling_summary_power_law} summarizes the scaling results.\\

\begin{table}[h!]
\centering
  \begin{tabular}{|p{0.15\textwidth}|p{0.12\textwidth}|p{0.12\textwidth}|}
  \hline
    & $\beta \sim O(1)$ & $\beta \gg 1$\\
  \hline
    Non-dimensional peeling time ($t_f$) & $\beta $ & $\beta^{-\frac{(1+n)}{n}} $ \\
    \hline
    Non-dimensional front width ($x_f$) & $ \left( t/t_f \right)^{ \frac{n}{n+1} } $ & $ \left( t/t_f \right)^{ \frac{n}{n+1} } $\\
        \hline
     Dimensional peeling time ($t_f^*$) & $ \frac{t_f}{{\delta}} \left( \frac{\Delta P^* {\delta}}{K^*} \right)^{-\frac{1}{n} } $ & $\frac{t_f}{{\delta}} \left( \frac{\Delta P^* {\delta}}{K^*} \right)^{-\frac{1}{n} } $ \\
    \hline
    Dimensional front width ($x_f^*$) & $ l^* \left( t^*/t_f^* \right)^{ \frac{n}{n+1} }$ & $ l^* \left( t^*/t_f^* \right)^{ \frac{n}{n+1} }$ \\
    \hline
\end{tabular}%
  \caption{Scaling relationships for front width and peeling time for a power law fluid with moderate channel deformations ($\beta \sim O(1)$ or smaller) and strong channel deformations ($\beta \gg 1$). The definitions of the FSI parameter $\beta$ and non-dimensional time are given by Eq. \eqref{powerlawbeta} and Eq. \eqref{powerlawtime}}
  \vspace{-1em}
\label{tbl:scaling_summary_power_law}
\end{table}

\underline{Moderate deformations ($\beta \sim O(1)$ or smaller)}:  For moderate conduit deformations, the above PDE admits a similarity solution at early times $t \ll \beta$.  The profile propagates as a front with thickness scaling as $x_f \sim \left( t/ \beta \right)^{\frac{n}{n+1}}$.  The front reaches the end when $x_f \sim O(1)$, which gives the peeling time as $t_f \sim \beta$. The form of the similarity solution is as follows:\\

\begin{equation}
    p = f(\eta);   \qquad \eta \equiv x \left(\frac{(2n + 1) \beta}{(n+1)t} \right)^{\frac{n}{n+1}}
\end{equation}

The above transformation converts the PDE \eqref{height_PL_evolution_substituted} into an ODE:

\begin{subequations}
\begin{align}
    &-\frac{d}{d \eta}\left( h^{\frac{2n+1}{n}} \left(-\frac{df}{d\eta}\right)^{\frac{1}{n}} \right) + \eta \frac{df}{d\eta} = 0;   \qquad h = 1 + \beta f \\
    &f(0) = 1; \quad f(\infty) = 0    
\end{align}
\end{subequations}

\begin{figure}
\centering
\subfloat{
  \includegraphics[width=0.5\textwidth]{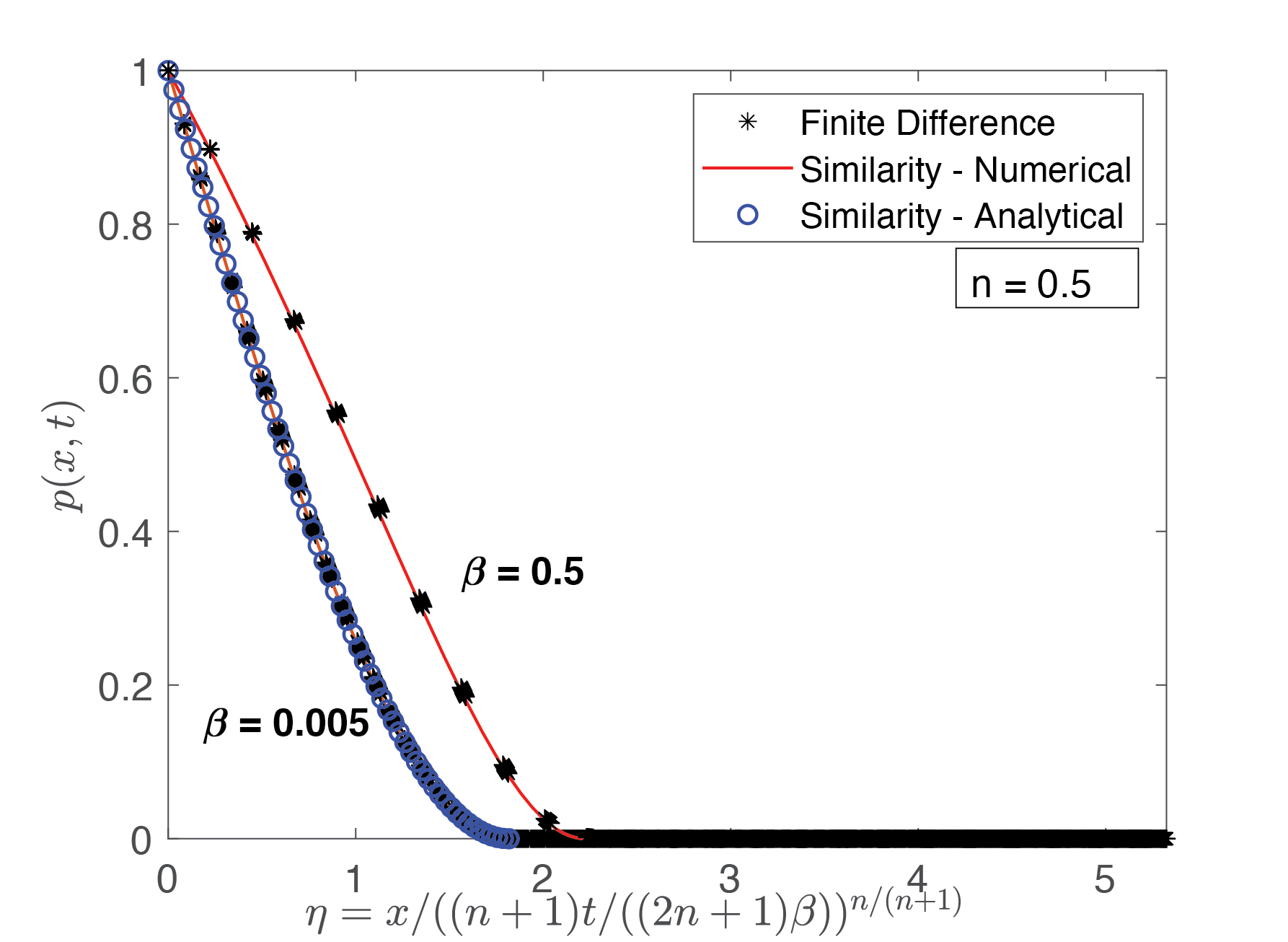}
}
\vspace{1em}
\subfloat{
  \includegraphics[width=0.5\textwidth]{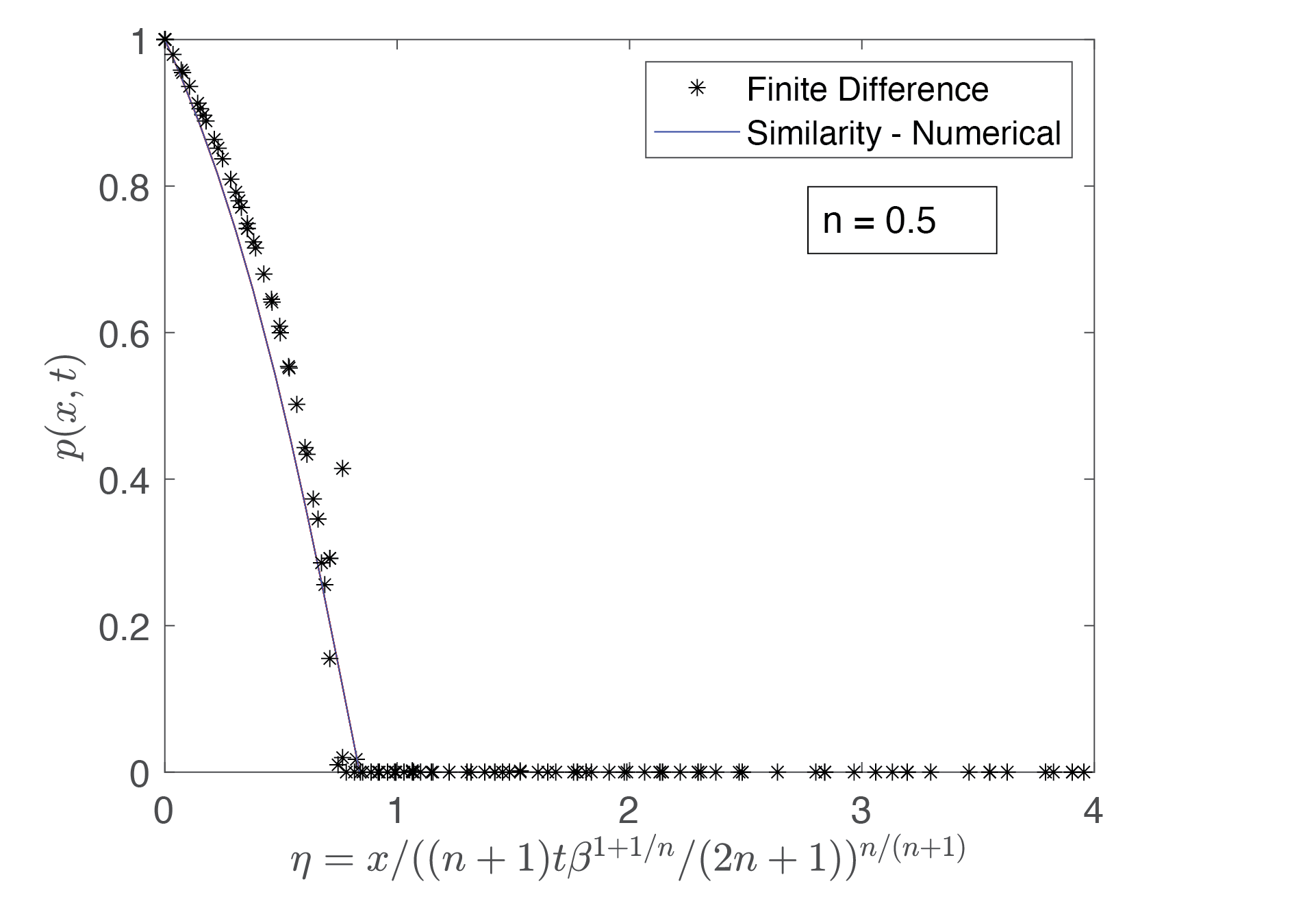}
}
\caption{Comparison between similarity solution and numerical solution for power law fluid.  (a) Moderate conduit deformations ($\beta \sim O(1)$ or smaller).  For $\beta = 0.005$, the time instances plotted for the numerical solution are $t = 10^{-7}, 10^{-5}$, and $5 \times 10^{-4}$.  The analytical solution is the leading order solution for $\beta \ll 1$ as stated in eqn (\ref{anal_soln_power_law}).  For $\beta = 0.5$, the time instances plotted for the numerical solution are $t = 10^{-7}, 10^{-5}$, and $1.5 \times 10^{-4}$.  (b) Strong conduit deformation ($\beta \gg 1$).  The plot is for $\beta = 10$ and the numerical solutions are plotted at time instants $t = 5 \times 10^{-7}, 5 \times 10^{-6}, 10^{-5}$, and $ 10^{-4}$.  For all plots, the shear-thinning index is $n = 0.5$.  The definitions of the FSI parameter $\beta$ and non-dimensional time are given by Eq. \eqref{powerlawbeta} and Eq. \eqref{powerlawtime}.  }
\label{PowerlawSimilarity}
\end{figure}

Again, just like in Section \ref{sec:moderate_sptt}, the above ODE is well defined only in a finite region $\eta \in [0, \eta_{max}]$. At a particular location $\eta = \eta_{max}$, the pressure gradient will equal zero and hence other physics will become important.  In actuality, the region around $\eta_{max}$ has a pressure so small that the power-law constitutive relationship likely no longer holds.  Thus, we consider replacing the boundary conditions of the ODE with the following:

\begin{equation}
f(0) = 1; \quad f(\eta_{max}) = 0; \qquad f'(\eta_{max}) = 0    
\end{equation}
where $\eta_{max}$ needs to be found.

For $\beta \ll 1$, the above similarity ODE admits an analytical solution:

\begin{subequations} \label{anal_soln_power_law}
    \begin{align}
        &f = 1 - \left( \frac{1-n}{2} \eta_{max}^2 \right)^{\frac{n}{1-n}} \eta_{max} \int_0^{\eta/\eta_{max}} \left[ 1 - z^2 \right]^{\frac{n}{1-n}}dz \\
        &\eta_{max}^{\frac{n+1}{1-n}} = \frac{ \left( \frac{2}{1-n} \right)^{\frac{n}{1-n}} }{ \int_0^1 \left( 1 - z^2 \right)^{ \frac{n}{1-n} } dz }
    \end{align}
\end{subequations}
    
For $\beta \sim O(1)$, we solve the ODE using the nested shooting method discussed before. The results have been shown in Fig.\ref{PowerlawSimilarity}.

From Table \ref{tbl:scaling_summary_power_law}, we note that the front propagation scaling - $x_f \sim t^{\frac{n}{n+1}}$ matches with the scaling law provided by Boyko et.al. \cite{boykogat2017} thereby suggesting that we could expect the analysis for power law fluids to also extend to long, thin walled, elastic shells. This can be owed to the fact that although the axisymmetric governing equations in both the problems differ by the addition of circumferential stresses in the tube case, the radially averaged model predicts a spring like behavior for the pressure drop-deformation relation. \\

\underline{Large deformation ($\beta \gg 1$)}:  When the conduit deformation is large $(\beta \gg 1)$, the conduit height behaves as $h \sim \beta \gg 1$, which gives rise to different scaling for the front width and peeling time.  For times $t \ll \beta^{-(1+n)/n}$ one obtains the scalings in Table \ref{tbl:scaling_summary2}.  The similarity transformation takes the following form:

\begin{equation}
    p = f(\eta);   \qquad \eta \equiv \frac{x}{\beta} \left(\frac{(2n + 1) }{(n+1)t} \right)^{\frac{n}{n+1}}
\end{equation}

The above transformation converts the PDE \eqref{height_PL_evolution_substituted} into an ODE:

\begin{subequations}
\begin{align}
    &-\frac{d}{d \eta}\left( \tilde{h}^{\frac{2n+1}{n}} \left(-\frac{df}{d\eta}\right)^{\frac{1}{n}} \right) + \eta \frac{df}{d\eta} = 0;   \qquad \tilde{h} = f + \beta^{-1} \\
    &f(0) = 1; \quad f(\infty) = 0    
\end{align}
\end{subequations}

Like the previous section, the equations are not well defined in the entire domain.  We will replace the boundary conditions with the following:

\begin{equation}
    f(1) = 0; \qquad f(\eta_{max}) = 0
\end{equation}
where $\eta_{max}$ is determined from simulation.

Fig.\ref{PowerlawSimilarity} shows the similarity solution for $\beta = 10$ and compares it with the numerical solution at different time instances.  Overall, the simulation data strongly suggests that similarity holds in the small time limit, with the solution given by the above ODE.

\section{Conclusion}\label{conclusion}

In this study, we explore a particular class of interactions between deformable slender sheets and viscoelastic fluids.  When two elastic sheets are separated by a lubricating layer of fluid, the flow of the fluid can cause the sheets to peel.  This situation is common in many applications in nature and industry, for example  exfoliation of graphene sheets\cite{Yi2016GrapheneReview}, hydraulic fracturing \cite{peng2020}, and cell sorting \cite{FuCytometry}. This study investigates how non-Newtonian liquids can alter peeling behavior of the sheets.  Specifically, the study quantifies the startup, transient behavior of the peeling process, characterizing how the peeling front propagates as well as the peeling time.  Two classes of complex fluids are studied – a simplified Phan-Thien Tanner (sPTT) fluid and a generalized power law fluid.  The sPTT model captures many of the qualitative features of polymeric solutions such as shear thinning, extensional thickening, viscoelasticity, and normal stress differences.  The second fluid only exhibits shear thinning, but can demonstrate a much wider range of shear thinning behavior than the sPTT model.

Our study finds that shear thinning plays a major role in modifying the peeling characteristics of the elastic sheet{. Other aspects of non Newtonian rheology including nonzero normal stress difference, elasticity, and nonlinearity are subdued for the system under consideration. To explain this observation for the s-PTT model we note that $\tau_{yy} = 0$ (see Eq.~\eqref{tau_{yy}} due to  the lubrication approximation, which makes the hydrodynamic normal load on the structure arise only from  pressure and not due to the deviatoric stress. On the other hand, because of nonlinearity of the rheological model, we have a finite $\tau_{xx}$ (see Eq.~\eqref{eq:normal_shear_relation}, even though the corresponding strain rate $\frac{\partial u}{\partial x} = 0$. This normal stress works to augment the shear thinning property of the fluid, as shown in Eq.\eqref{eq:tau_xyDe}, and ultimately alters the lubrication pressure. Finally, the assumptions pertaining to small $\epsilon Wi $ allow us to neglect the elasticity/relaxation time of the fluid completely.}  

When compared to a Newtonian fluid with equivalent zero-shear viscosity, a shear thinning liquid will have its peeling front propagate with a smaller power law exponent than a Newtonian fluid (e.g., $x_f \sim t^{-n/(n+1)}$ for a power law index $n$, compared to $x_f \sim t^{1/2}$ for Newtonian), but with an order-of-magnitude larger prefactor.  The overall consequence of this result is that the peeling time for the shear thinning fluid is much smaller than the equivalent Newtonian fluid.  For both constitutive relationships studied, scaling relationships are provided for the peeling front and peeling time in the Newtonian and non-Newtonian (e.g., power law) regions of the viscosity curves, under the specific limits of moderate sheet deformation and strong sheet deformation (see Tables \ref{tbl:powerlaw}-\ref{tbl:scaling_summary2}).  Similarity solutions are provided to give a quantitative description in these limiting regimes, as well as full numerical solutions in the regimes where the similarity solutions no longer hold (e.g., when ${\varepsilon}Wi^2 \sim O(1)$ for the sPTT fluid, where $Wi$ is the Weissenberg number and ${\varepsilon}$ is the elongation parameter).  We hope that these results will give much needed insight for those interested in manipulating slender, deformable sheet structures with non-Newtonian liquids.

\section{Acknowledgements}
The authors acknowledge funding from the American Chemical Society Petroleum Research Fund (Grant No. ACS PRF 61266-DNI9).

\appendix
\section{Derivation of flow controlled case solution at steady state} \label{app:flowratecontrolled}
From section \ref{Height}, we know that the height evolution equation is given as follows:

\begin{equation}
    {q} = -\left(\frac{d{p}}{d{x}}\right)\left(\frac{2h^3}{3}\right) - \left(\frac{d{p}}{d{x}}\right)^3\left(\frac{{\varepsilon} Wi^2}{5}\right)h^5
\end{equation}

If we set the non-dimensional flow rate as $1$, we get the following governing equation:

\begin{equation}
\label{Flow Rate Controlled GovEqn}
    1 = -\left(\frac{d{p}}{d{x}}\right)\left(\frac{2h^3}{3}\right) - \left(\frac{d {p}}{d{x}}\right)^3\left(\frac{{\varepsilon} Wi^2}{5}\right)h^5
\end{equation}

From the pressure-deformation relation, we have 
\begin{equation}
\label{Pressure-Deformation_Final}
    {h} = 1 + \beta{p}
\end{equation}

which when differentiated, yields 
\begin{equation}
\label{Pressure-Deformation_Diff}
    \frac{d{h}}{d{x}} = \beta\frac{d{p}}{d{x}} \implies  \frac{1}{\beta}\frac{d{h}}{d{x}} = \frac{d{p}}{d{x}}
\end{equation}

On substituting \eqref{Pressure-Deformation_Diff} into \eqref{Flow Rate Controlled GovEqn}, we get 

\begin{equation}
\label{Height_FRC}
    1  + \frac{1}{\beta}\left(\frac{dh}{dx}\right)\left(\frac{2h^3}{3}\right) + \frac{1}{\beta^{3}}\left(\frac{dh}{dx}\right)^3\left(\frac{{\varepsilon} Wi^2}{5}\right)h^5 = 0
\end{equation}

We make the following transformation: 
\begin{equation}
{h^{2}} = {t_{1}} \implies {2h}\frac{d{h}}{d{x}} = \frac{d{t_{1}}}{d{x}}
\end{equation}

This transformation converts \eqref{Height_FRC} into

 {\begin{equation}
\left(\frac{d{t_{1}}}{d{x}} \right)^3\left(\frac{t_{1}kWi^{2}}{40\beta^{3}}\right) + \left(\frac{d{t_{1}}}{d{x}}\right)\left(\frac{t_{1}}{3\beta}\right) + 1 = 0
\end{equation}
}

\begin{equation}
\frac{d{t_{1}}}{d{x}} = \theta
\end{equation}

On choosing the variables  \large${a} = \frac{t_{1}{\varepsilon} Wi^{2}}{40\beta^{3}} $\large ;  \large$ {b} = \frac{t_{1}}{3\beta} $\large ;   \large${c} = 1$\large

\normalsize We get the following depressed cubic equation:

\begin{equation}
    a\theta^{3} + b\theta + c = 0
\end{equation}

If we write the terms as \large$\frac{b}{a} = {p_1}$ { }{ };{ }{ } \large$\frac{c}{a} = {q_1}$

\normalsize This equation has a well known solution given by the Cardano's formula:

\begin{equation}
    \theta = \left(-\frac{q_1}{2} + \left(\frac{q_{1}^2}{4} + \frac{p_{1}^3}{27}\right)^\frac{1}{2}\right)^{\frac{1}{3}} + \left(-\frac{q_{1}}{2} - \left(\frac{q_{1}^2}{4} + \frac{p_{1}^3}{27}\right)^\frac{1}{2}\right)^{\frac{1}{3}}
\end{equation}

This creates the ordinary differential 
\begin{equation}
    \frac{d{t_{1}}}{d{x}} = \left(-\frac{q}{2} + \left(\frac{q^2}{4} + \frac{p^3}{27}\right)^\frac{1}{2}\right)^{\frac{1}{3}} + \left(-\frac{q}{2} - \left(\frac{q^2}{4} + \frac{p^3}{27}\right)^\frac{1}{2}\right)^{\frac{1}{3}}
\end{equation}

We perform a simple transformation \large$p = m$ and \large$q = \frac{n}{t_{1}}$

\normalsize{\begin{equation}
 \frac{d{t_{1}}}{d{x}} = \left(-\frac{n}{2t_{1}} + \left(\frac{n^2}{4t_{1}^2} + \frac{m^3}{27}\right)^\frac{1}{2}\right)^{\frac{1}{3}} + \left(-\frac{n}{2t_{1}} - \left(\frac{n^2}{4t_{1}^2} + \frac{m^3}{27}\right)^\frac{1}{2}\right)^{\frac{1}{3}}
\end{equation}}

We now take \large $M = \frac{n}{2m^{3/2}} = \left(\frac{27\textcolor{blue}{\varepsilon} Wi^{2}}{160}\right)^{1/2} $ 
\normalsize{\begin{multline}
    \frac{d{t_{1}}}{d{x}} = m^{1/2}\Big(\left(-\frac{M}{t_{1}} + \left( \frac{M^{2}}{t_{1}^{2}} + \frac{1}{27}\right)^{1/2}\right)^{1/3} +\\ \left(-\frac{M}{t_{1}} - \left( \frac{M^{2}}{t_{1}^{2}} + \frac{1}{27}\right)^{1/2}\right)^{1/3} \Big)
\end{multline}}
On substituting $t_{1} = 3\sqrt{3}M \tan\theta$, we get the following equation
\begin{equation}
    \frac{9M}{\sqrt{m}} \frac{\sec^{2}\theta \tan^{1/3}\theta}{(\sec\theta - 1)^{1/3} - (\sec\theta + 1)^{1/3}}d\theta = dx
\end{equation}
This can be further simplified to give us :
\begin{equation}
    \frac{9M}{\sqrt{m}} \frac{\sec^{2}\theta \tan^{1/3}\frac{\theta}{2}}{\tan^{2/3}\frac{\theta}{2} - 1}d\theta  = dx
\end{equation}
We then take $v = (\tan\frac{\theta}{2})^{1/3}$  
\begin{equation}
     \frac{54M}{\sqrt{m}}\frac{(v^6 + 1)(v^3) dv}{(v^2 - 1)(v^6 - 1)^2} = dx
\end{equation}
Using partial fractions, we can get the final form of the integrated equation
\begin{equation}
    I(v) = x + C_{1}
\end{equation}
where $I(v)$ is given by :
\normalsize{\begin{equation}
    I(v)  = \frac{54M}{\sqrt{m}}\left[ \frac{log|v^2 - 1|}{27} - \frac{log|v^4 + v^2 + 1|}{54} + \frac{v^6 - 5v^4 + v^2}{18(v^2 - 1)(v^6 - 1)} \right]
\end{equation}}
and $C_{1}$ is an integration constant. Furthermore, on substituting the BC $h(x=1) = 1$
we get an implicit equation for $h(x)$ and $x$ given by :
\begin{equation}
    I(v) = x + I(v_1) 
\end{equation}
where $v_1 = (\tan\frac{\theta_1}{2})^{1/3}$ and $\theta_1 = \arctan{\frac{1}{3\sqrt{3}M}}$

\section{ Equations for finite difference }

We use a central difference scheme for the spatial derivatives given by :

\begin{equation}
    \frac{\partial{p}}{\partial{x}} \equiv \frac{p^{n}_{i+1} - 2p^{n}_{i} + p^{n}_{i-1}}{(\Delta x)^2} 
\end{equation}
and temporal derivatives using the backward difference scheme :
\begin{equation}
    \frac{\partial p}{\partial t} \equiv \frac{p^{n}_i - p^{n-1}_i}{\Delta t}
\end{equation}

The value of pressure at the $(n-1)^{th}$ step is taken to be the same as the $k^{th}$ Newton iteration of the $(n-1)^{th}$ step. The discretized nonlinear equation, whose Jacobian is calculated for the Newton iterations, is given as follows:

\normalsize{\begin{multline}
    \frac{p^{n}_i - p^{n-1}_i}{\Delta t} - \frac{2}{3\beta}\left(\frac{p^{n}_{i+1} - 2p^{n}_{i} + p^{n}_{i-1}}{(\Delta x)^2}\right) - 2(1+\beta p^{n}_i)^{2}\left(\frac{p^{n}_{i+1} -  p^{n}_{i-1}}{2\Delta x}\right)^{2} \\- \left(\frac{6{\varepsilon}Wi^{2}}{5\beta}\right)\left(\frac{p^{n}_{i+1} -  p^{n}_{i-1}}{2\Delta x}\right)^{2}\left(\frac{p^{n}_{i+1} - 2p^{n}_{i} + p^{n}_{i-1}}{(\Delta x)^2}\right)(1+\beta p^n_{i})^{5} -\\ \left(\frac{2{\varepsilon}Wi^{2}}{5}\right)\left(\frac{p^{n}_{i+1} -  p^{n}_{i-1}}{2\Delta x}\right)^{4}(1+\beta p^n_{i})^{4} \equiv f(p^n_{i+1},p^n_{i},p^n_{i-1})
\end{multline}}

We then proceed to calculate the Jacobian of $f$ and solve the Newton's method problem until an error of $1e-7$ is attained based on the absolute value of the difference between $p^{k}$ and $p^{k+1}$.


\bibliographystyle{elsarticle-num.bst}
\bibliography{refs.bib}


\end{document}